\magnification=\magstep1
\hsize=31pc 
\vsize=49pc 
\lineskip=0pt 
\parskip=0pt plus 1pt 
\hfuzz=1pt   
\vfuzz=2pt 
\pretolerance=2500 
\tolerance=5000 
\vbadness=5000 
\hbadness=5000 
\widowpenalty=500 
\clubpenalty=200 
\brokenpenalty=500 
\predisplaypenalty=200 
\voffset=-1pc 
\nopagenumbers      
\catcode`@=11 
\newif\ifams 
\amsfalse 
%
%
%
\newfam\bdifam 
\newfam\bsyfam 
\newfam\bssfam 
\newfam\msafam 
\newfam\msbfam 
\newif\ifxxpt    
\newif\ifxviipt  
\newif\ifxivpt   
\newif\ifxiipt   
\newif\ifxipt    
\newif\ifxpt     
\newif\ifixpt    
\newif\ifviiipt  
\newif\ifviipt   
\newif\ifvipt    
\newif\ifvpt     
%
%
\def\headsize#1#2{\def\headb@seline{#2}%
                \ifnum#1=20\def\HEAD{twenty}%
                           \def\smHEAD{twelve}%
                           \def\vsHEAD{nine}%
                           \ifxxpt\else\xdef\f@ntsize{\HEAD}%
                           \def\m@g{4}\def\s@ze{20.74}%
                           \loadheadfonts\xxpttrue\fi 
                           \ifxiipt\else\xdef\f@ntsize{\smHEAD}%
                           \def\m@g{1}\def\s@ze{12}%
                           \loadxiiptfonts\xiipttrue\fi 
                           \ifixpt\else\xdef\f@ntsize{\vsHEAD}%
                           \def\s@ze{9}%
                           \loadsmallfonts\ixpttrue\fi 
                      \else 
                \ifnum#1=17\def\HEAD{seventeen}%
                           \def\smHEAD{eleven}%
                           \def\vsHEAD{eight}%
                           \ifxviipt\else\xdef\f@ntsize{\HEAD}%
                           \def\m@g{3}\def\s@ze{17.28}%
                           \loadheadfonts\xviipttrue\fi 
                           \ifxipt\else\xdef\f@ntsize{\smHEAD}%
                           \loadxiptfonts\xipttrue\fi 
                           \ifviiipt\else\xdef\f@ntsize{\vsHEAD}%
                           \def\s@ze{8}%
                           \loadsmallfonts\viiipttrue\fi 
                      \else\def\HEAD{fourteen}%
                           \def\smHEAD{ten}%
                           \def\vsHEAD{seven}%
                           \ifxivpt\else\xdef\f@ntsize{\HEAD}%
                           \def\m@g{2}\def\s@ze{14.4}%
                           \loadheadfonts\xivpttrue\fi 
                           \ifxpt\else\xdef\f@ntsize{\smHEAD}%
                           \def\s@ze{10}%
                           \loadxptfonts\xpttrue\fi 
                           \ifviipt\else\xdef\f@ntsize{\vsHEAD}%
                           \def\s@ze{7}%
                           \loadviiptfonts\viipttrue\fi 
                \ifnum#1=14\else 
                \message{Header size should be 20, 17 or 14 point 
                              will now default to 14pt}\fi 
                \fi\fi\headfonts} 
%
%
\def\textsize#1#2{\def\textb@seline{#2}%
                 \ifnum#1=12\def\TEXT{twelve}%
                           \def\smTEXT{eight}%
                           \def\vsTEXT{six}%
                           \ifxiipt\else\xdef\f@ntsize{\TEXT}%
                           \def\m@g{1}\def\s@ze{12}%
                           \loadxiiptfonts\xiipttrue\fi 
                           \ifviiipt\else\xdef\f@ntsize{\smTEXT}%
                           \def\s@ze{8}%
                           \loadsmallfonts\viiipttrue\fi 
                           \ifvipt\else\xdef\f@ntsize{\vsTEXT}%
                           \def\s@ze{6}%
                           \loadviptfonts\vipttrue\fi 
                      \else 
                \ifnum#1=11\def\TEXT{eleven}%
                           \def\smTEXT{seven}%
                           \def\vsTEXT{five}%
                           \ifxipt\else\xdef\f@ntsize{\TEXT}%
                           \def\s@ze{11}%
                           \loadxiptfonts\xipttrue\fi 
                           \ifviipt\else\xdef\f@ntsize{\smTEXT}%
                           \loadviiptfonts\viipttrue\fi 
                           \ifvpt\else\xdef\f@ntsize{\vsTEXT}%
                           \def\s@ze{5}%
                           \loadvptfonts\vpttrue\fi 
                      \else\def\TEXT{ten}%
                           \def\smTEXT{seven}%
                           \def\vsTEXT{five}%
                           \ifxpt\else\xdef\f@ntsize{\TEXT}%
                           \loadxptfonts\xpttrue\fi 
                           \ifviipt\else\xdef\f@ntsize{\smTEXT}%
                           \def\s@ze{7}%
                           \loadviiptfonts\viipttrue\fi 
                           \ifvpt\else\xdef\f@ntsize{\vsTEXT}%
                           \def\s@ze{5}%
                           \loadvptfonts\vpttrue\fi 
                \ifnum#1=10\else 
                \message{Text size should be 12, 11 or 10 point 
                              will now default to 10pt}\fi 
                \fi\fi\textfonts} 
%
%
\def\smallsize#1#2{\def\smallb@seline{#2}%
                 \ifnum#1=10\def\SMALL{ten}%
                           \def\smSMALL{seven}%
                           \def\vsSMALL{five}%
                           \ifxpt\else\xdef\f@ntsize{\SMALL}%
                           \loadxptfonts\xpttrue\fi 
                           \ifviipt\else\xdef\f@ntsize{\smSMALL}%
                           \def\s@ze{7}%
                           \loadviiptfonts\viipttrue\fi 
                           \ifvpt\else\xdef\f@ntsize{\vsSMALL}%
                           \def\s@ze{5}%
                           \loadvptfonts\vpttrue\fi 
                       \else 
                 \ifnum#1=9\def\SMALL{nine}%
                           \def\smSMALL{six}%
                           \def\vsSMALL{five}%
                           \ifixpt\else\xdef\f@ntsize{\SMALL}%
                           \def\s@ze{9}%
                           \loadsmallfonts\ixpttrue\fi 
                           \ifvipt\else\xdef\f@ntsize{\smSMALL}%
                           \def\s@ze{6}%
                           \loadviptfonts\vipttrue\fi 
                           \ifvpt\else\xdef\f@ntsize{\vsSMALL}%
                           \def\s@ze{5}%
                           \loadvptfonts\vpttrue\fi 
                       \else 
                           \def\SMALL{eight}%
                           \def\smSMALL{six}%
                           \def\vsSMALL{five}%
                           \ifviiipt\else\xdef\f@ntsize{\SMALL}%
                           \def\s@ze{8}%
                           \loadsmallfonts\viiipttrue\fi 
                           \ifvipt\else\xdef\f@ntsize{\smSMALL}%
                           \def\s@ze{6}%
                           \loadviptfonts\vipttrue\fi 
                           \ifvpt\else\xdef\f@ntsize{\vsSMALL}%
                           \def\s@ze{5}%
                           \loadvptfonts\vpttrue\fi 
                 \ifnum#1=8\else\message{Small size should be 10, 9 or  
                            8 point will now default to 8pt}\fi 
                \fi\fi\smallfonts} 
\def\F@nt{\expandafter\font\csname} 
\def\Sk@w{\expandafter\skewchar\csname} 
\def\@nd{\endcsname} 
\def\@step#1{ scaled \magstep#1} 
\def\@half{ scaled \magstephalf} 
\def\@t#1{ at #1pt} 
%
%
\def\loadheadfonts{\bigf@nts 
\F@nt \f@ntsize bdi\@nd=cmmib10 \@t{\s@ze}%
\Sk@w \f@ntsize bdi\@nd='177 
\F@nt \f@ntsize bsy\@nd=cmbsy10 \@t{\s@ze}%
\Sk@w \f@ntsize bsy\@nd='60 
\F@nt \f@ntsize bss\@nd=cmssbx10 \@t{\s@ze}} 
%
%
\def\loadxiiptfonts{\bigf@nts 
\F@nt \f@ntsize bdi\@nd=cmmib10 \@step{\m@g}%
\Sk@w \f@ntsize bdi\@nd='177 
\F@nt \f@ntsize bsy\@nd=cmbsy10 \@step{\m@g}%
\Sk@w \f@ntsize bsy\@nd='60 
\F@nt \f@ntsize bss\@nd=cmssbx10 \@step{\m@g}} 
%
%
\def\loadxiptfonts{%
\font\elevenrm=cmr10 \@half 
\font\eleveni=cmmi10 \@half 
\skewchar\eleveni='177 
\font\elevensy=cmsy10 \@half 
\skewchar\elevensy='60 
\font\elevenex=cmex10 \@half 
\font\elevenit=cmti10 \@half 
\font\elevensl=cmsl10 \@half 
\font\elevenbf=cmbx10 \@half 
\font\eleventt=cmtt10 \@half 
\ifams\font\elevenmsa=msam10 \@half 
\font\elevenmsb=msbm10 \@half\else\fi 
\font\elevenbdi=cmmib10 \@half 
\skewchar\elevenbdi='177 
\font\elevenbsy=cmbsy10 \@half 
\skewchar\elevenbsy='60 
\font\elevenbss=cmssbx10 \@half} 
%
%
\def\loadxptfonts{%
\font\tenbdi=cmmib10 
\skewchar\tenbdi='177 
\font\tenbsy=cmbsy10  
\skewchar\tenbsy='60 
\ifams\font\tenmsa=msam10  
\font\tenmsb=msbm10\else\fi 
\font\tenbss=cmssbx10}%
%
%
\def\loadsmallfonts{\smallf@nts 
\ifams 
\F@nt \f@ntsize ex\@nd=cmex\s@ze 
\else 
\F@nt \f@ntsize ex\@nd=cmex10\fi 
\F@nt \f@ntsize it\@nd=cmti\s@ze 
\F@nt \f@ntsize sl\@nd=cmsl\s@ze 
\F@nt \f@ntsize tt\@nd=cmtt\s@ze} 
%
%
\def\loadviiptfonts{%
\font\sevenit=cmti7 
\font\sevensl=cmsl8 at 7pt 
\ifams\font\sevenmsa=msam7  
\font\sevenmsb=msbm7 
\font\sevenex=cmex7 
\font\sevenbsy=cmbsy7 
\font\sevenbdi=cmmib7\else 
\font\sevenex=cmex10 
\font\sevenbsy=cmbsy10 at 7pt 
\font\sevenbdi=cmmib10 at 7pt\fi 
\skewchar\sevenbsy='60 
\skewchar\sevenbdi='177 
\font\sevenbss=cmssbx10 at 7pt}%
%
%
\def\loadviptfonts{\smallf@nts 
\ifams\font\sixex=cmex7 at 6pt\else 
\font\sixex=cmex10\fi 
\font\sixit=cmti7 at 6pt} 
%
%
\def\loadvptfonts{%
\font\fiveit=cmti7 at 5pt 
\ifams\font\fiveex=cmex7 at 5pt 
\font\fivebdi=cmmib5 
\font\fivebsy=cmbsy5 
\font\fivemsa=msam5  
\font\fivemsb=msbm5\else 
\font\fiveex=cmex10 
\font\fivebdi=cmmib10 at 5pt 
\font\fivebsy=cmbsy10 at 5pt\fi 
\skewchar\fivebdi='177 
\skewchar\fivebsy='60 
\font\fivebss=cmssbx10 at 5pt} 
\def\bigf@nts{%
\F@nt \f@ntsize rm\@nd=cmr10 \@step{\m@g}%
\F@nt \f@ntsize i\@nd=cmmi10 \@step{\m@g}%
\Sk@w \f@ntsize i\@nd='177 
\F@nt \f@ntsize sy\@nd=cmsy10 \@step{\m@g}%
\Sk@w \f@ntsize sy\@nd='60 
\F@nt \f@ntsize ex\@nd=cmex10 \@step{\m@g}%
\F@nt \f@ntsize it\@nd=cmti10 \@step{\m@g}%
\F@nt \f@ntsize sl\@nd=cmsl10 \@step{\m@g}%
\F@nt \f@ntsize bf\@nd=cmbx10 \@step{\m@g}%
\F@nt \f@ntsize tt\@nd=cmtt10 \@step{\m@g}%
\ifams 
\F@nt \f@ntsize msa\@nd=msam10 \@step{\m@g}%
\F@nt \f@ntsize msb\@nd=msbm10 \@step{\m@g}\else\fi} 
\def\smallf@nts{%
\F@nt \f@ntsize rm\@nd=cmr\s@ze 
\F@nt \f@ntsize i\@nd=cmmi\s@ze  
\Sk@w \f@ntsize i\@nd='177 
\F@nt \f@ntsize sy\@nd=cmsy\s@ze 
\Sk@w \f@ntsize sy\@nd='60 
\F@nt \f@ntsize bf\@nd=cmbx\s@ze  
\ifams 
\F@nt \f@ntsize bdi\@nd=cmmib\s@ze  
\F@nt \f@ntsize bsy\@nd=cmbsy\s@ze  
\F@nt \f@ntsize msa\@nd=msam\s@ze  
\F@nt \f@ntsize msb\@nd=msbm\s@ze 
\else 
\F@nt \f@ntsize bdi\@nd=cmmib10 \@t{\s@ze}%
\F@nt \f@ntsize bsy\@nd=cmbsy10 \@t{\s@ze}\fi  
\Sk@w \f@ntsize bdi\@nd='177 
\Sk@w \f@ntsize bsy\@nd='60 
\F@nt \f@ntsize bss\@nd=cmssbx10 \@t{\s@ze}}%
%
%
\def\headfonts{%
\textfont0=\csname\HEAD rm\@nd         
\scriptfont0=\csname\smHEAD rm\@nd 
\scriptscriptfont0=\csname\vsHEAD rm\@nd 
\def\rm{\fam0\csname\HEAD rm\@nd 
\def\sc{\csname\smHEAD rm\@nd}}%
\textfont1=\csname\HEAD i\@nd          
\scriptfont1=\csname\smHEAD i\@nd 
\scriptscriptfont1=\csname\vsHEAD i\@nd 
\textfont2=\csname\HEAD sy\@nd         
\scriptfont2=\csname\smHEAD sy\@nd 
\scriptscriptfont2=\csname\vsHEAD sy\@nd 
\textfont3=\csname\HEAD ex\@nd         
\scriptfont3=\csname\smHEAD ex\@nd 
\scriptscriptfont3=\csname\smHEAD ex\@nd 
\textfont\itfam=\csname\HEAD it\@nd    
\scriptfont\itfam=\csname\smHEAD it\@nd 
\scriptscriptfont\itfam=\csname\vsHEAD it\@nd 
\def\it{\fam\itfam\csname\HEAD it\@nd 
\def\sc{\csname\smHEAD it\@nd}}%
\textfont\slfam=\csname\HEAD sl\@nd    
\def\sl{\fam\slfam\csname\HEAD sl\@nd 
\def\sc{\csname\smHEAD sl\@nd}}%
\textfont\bffam=\csname\HEAD bf\@nd    
\scriptfont\bffam=\csname\smHEAD bf\@nd 
\scriptscriptfont\bffam=\csname\vsHEAD bf\@nd 
\def\bf{\fam\bffam\csname\HEAD bf\@nd 
\def\sc{\csname\smHEAD bf\@nd}}%
\textfont\ttfam=\csname\HEAD tt\@nd    
\def\tt{\fam\ttfam\csname\HEAD tt\@nd}%
\textfont\bdifam=\csname\HEAD bdi\@nd  
\scriptfont\bdifam=\csname\smHEAD bdi\@nd 
\scriptscriptfont\bdifam=\csname\vsHEAD bdi\@nd 
\def\bdi{\fam\bdifam\csname\HEAD bdi\@nd}%
\textfont\bsyfam=\csname\HEAD bsy\@nd  
\scriptfont\bsyfam=\csname\smHEAD bsy\@nd 
\def\bsy{\fam\bsyfam\csname\HEAD bsy\@nd}%
\textfont\bssfam=\csname\HEAD bss\@nd  
\scriptfont\bssfam=\csname\smHEAD bss\@nd 
\scriptscriptfont\bssfam=\csname\vsHEAD bss\@nd 
\def\bss{\fam\bssfam\csname\HEAD bss\@nd}%
\ifams 
\textfont\msafam=\csname\HEAD msa\@nd  
\scriptfont\msafam=\csname\smHEAD msa\@nd 
\scriptscriptfont\msafam=\csname\vsHEAD msa\@nd 
\textfont\msbfam=\csname\HEAD msb\@nd  
\scriptfont\msbfam=\csname\smHEAD msb\@nd 
\scriptscriptfont\msbfam=\csname\vsHEAD msb\@nd 
\else\fi 
\normalbaselineskip=\headb@seline pt%
\setbox\strutbox=\hbox{\vrule height.7\normalbaselineskip  
depth.3\baselineskip width0pt}%
\def\sc{\csname\smHEAD rm\@nd}\normalbaselines\bf} 
%
%
\def\textfonts{%
\textfont0=\csname\TEXT rm\@nd         
\scriptfont0=\csname\smTEXT rm\@nd 
\scriptscriptfont0=\csname\vsTEXT rm\@nd 
\def\rm{\fam0\csname\TEXT rm\@nd 
\def\sc{\csname\smTEXT rm\@nd}}%
\textfont1=\csname\TEXT i\@nd          
\scriptfont1=\csname\smTEXT i\@nd 
\scriptscriptfont1=\csname\vsTEXT i\@nd 
\textfont2=\csname\TEXT sy\@nd         
\scriptfont2=\csname\smTEXT sy\@nd 
\scriptscriptfont2=\csname\vsTEXT sy\@nd 
\textfont3=\csname\TEXT ex\@nd         
\scriptfont3=\csname\smTEXT ex\@nd 
\scriptscriptfont3=\csname\smTEXT ex\@nd 
\textfont\itfam=\csname\TEXT it\@nd    
\scriptfont\itfam=\csname\smTEXT it\@nd 
\scriptscriptfont\itfam=\csname\vsTEXT it\@nd 
\def\it{\fam\itfam\csname\TEXT it\@nd 
\def\sc{\csname\smTEXT it\@nd}}%
\textfont\slfam=\csname\TEXT sl\@nd    
\def\sl{\fam\slfam\csname\TEXT sl\@nd 
\def\sc{\csname\smTEXT sl\@nd}}%
\textfont\bffam=\csname\TEXT bf\@nd    
\scriptfont\bffam=\csname\smTEXT bf\@nd 
\scriptscriptfont\bffam=\csname\vsTEXT bf\@nd 
\def\bf{\fam\bffam\csname\TEXT bf\@nd 
\def\sc{\csname\smTEXT bf\@nd}}%
\textfont\ttfam=\csname\TEXT tt\@nd    
\def\tt{\fam\ttfam\csname\TEXT tt\@nd}%
\textfont\bdifam=\csname\TEXT bdi\@nd  
\scriptfont\bdifam=\csname\smTEXT bdi\@nd 
\scriptscriptfont\bdifam=\csname\vsTEXT bdi\@nd 
\def\bdi{\fam\bdifam\csname\TEXT bdi\@nd}%
\textfont\bsyfam=\csname\TEXT bsy\@nd  
\scriptfont\bsyfam=\csname\smTEXT bsy\@nd 
\def\bsy{\fam\bsyfam\csname\TEXT bsy\@nd}%
\textfont\bssfam=\csname\TEXT bss\@nd  
\scriptfont\bssfam=\csname\smTEXT bss\@nd 
\scriptscriptfont\bssfam=\csname\vsTEXT bss\@nd 
\def\bss{\fam\bssfam\csname\TEXT bss\@nd}%
\ifams 
\textfont\msafam=\csname\TEXT msa\@nd  
\scriptfont\msafam=\csname\smTEXT msa\@nd 
\scriptscriptfont\msafam=\csname\vsTEXT msa\@nd 
\textfont\msbfam=\csname\TEXT msb\@nd  
\scriptfont\msbfam=\csname\smTEXT msb\@nd 
\scriptscriptfont\msbfam=\csname\vsTEXT msb\@nd 
\else\fi 
\normalbaselineskip=\textb@seline pt 
\setbox\strutbox=\hbox{\vrule height.7\normalbaselineskip  
depth.3\baselineskip width0pt}%
\everymath{}%
\def\sc{\csname\smTEXT rm\@nd}\normalbaselines\rm} 
%
%
\def\smallfonts{%
\textfont0=\csname\SMALL rm\@nd         
\scriptfont0=\csname\smSMALL rm\@nd 
\scriptscriptfont0=\csname\vsSMALL rm\@nd 
\def\rm{\fam0\csname\SMALL rm\@nd 
\def\sc{\csname\smSMALL rm\@nd}}%
\textfont1=\csname\SMALL i\@nd          
\scriptfont1=\csname\smSMALL i\@nd 
\scriptscriptfont1=\csname\vsSMALL i\@nd 
\textfont2=\csname\SMALL sy\@nd         
\scriptfont2=\csname\smSMALL sy\@nd 
\scriptscriptfont2=\csname\vsSMALL sy\@nd 
\textfont3=\csname\SMALL ex\@nd         
\scriptfont3=\csname\smSMALL ex\@nd 
\scriptscriptfont3=\csname\smSMALL ex\@nd 
\textfont\itfam=\csname\SMALL it\@nd    
\scriptfont\itfam=\csname\smSMALL it\@nd 
\scriptscriptfont\itfam=\csname\vsSMALL it\@nd 
\def\it{\fam\itfam\csname\SMALL it\@nd 
\def\sc{\csname\smSMALL it\@nd}}%
\textfont\slfam=\csname\SMALL sl\@nd    
\def\sl{\fam\slfam\csname\SMALL sl\@nd 
\def\sc{\csname\smSMALL sl\@nd}}%
\textfont\bffam=\csname\SMALL bf\@nd    
\scriptfont\bffam=\csname\smSMALL bf\@nd 
\scriptscriptfont\bffam=\csname\vsSMALL bf\@nd 
\def\bf{\fam\bffam\csname\SMALL bf\@nd 
\def\sc{\csname\smSMALL bf\@nd}}%
\textfont\ttfam=\csname\SMALL tt\@nd    
\def\tt{\fam\ttfam\csname\SMALL tt\@nd}%
\textfont\bdifam=\csname\SMALL bdi\@nd  
\scriptfont\bdifam=\csname\smSMALL bdi\@nd 
\scriptscriptfont\bdifam=\csname\vsSMALL bdi\@nd 
\def\bdi{\fam\bdifam\csname\SMALL bdi\@nd}%
\textfont\bsyfam=\csname\SMALL bsy\@nd  
\scriptfont\bsyfam=\csname\smSMALL bsy\@nd 
\def\bsy{\fam\bsyfam\csname\SMALL bsy\@nd}%
\textfont\bssfam=\csname\SMALL bss\@nd  
\scriptfont\bssfam=\csname\smSMALL bss\@nd 
\scriptscriptfont\bssfam=\csname\vsSMALL bss\@nd 
\def\bss{\fam\bssfam\csname\SMALL bss\@nd}%
\ifams 
\textfont\msafam=\csname\SMALL msa\@nd  
\scriptfont\msafam=\csname\smSMALL msa\@nd 
\scriptscriptfont\msafam=\csname\vsSMALL msa\@nd 
\textfont\msbfam=\csname\SMALL msb\@nd  
\scriptfont\msbfam=\csname\smSMALL msb\@nd 
\scriptscriptfont\msbfam=\csname\vsSMALL msb\@nd 
\else\fi 
\normalbaselineskip=\smallb@seline pt%
\setbox\strutbox=\hbox{\vrule height.7\normalbaselineskip  
depth.3\baselineskip width0pt}%
\everymath{}%
\def\sc{\csname\smSMALL rm\@nd}\normalbaselines\rm}%
\everydisplay{\indenteddisplay 
   \gdef\labeltype{\eqlabel}}%
%
%
\def\hexnumber@#1{\ifcase#1 0\or 1\or 2\or 3\or 4\or 5\or 6\or 7\or 8\or 
 9\or A\or B\or C\or D\or E\or F\fi} 
\edef\bffam@{\hexnumber@\bffam} 
\edef\bdifam@{\hexnumber@\bdifam} 
\edef\bsyfam@{\hexnumber@\bsyfam} 
\def\undefine#1{\let#1\undefined} 
\def\newsymbol#1#2#3#4#5{\let\next@\relax 
 \ifnum#2=\thr@@\let\next@\bdifam@\else 
 \ifams 
 \ifnum#2=\@ne\let\next@\msafam@\else 
 \ifnum#2=\tw@\let\next@\msbfam@\fi\fi 
 \fi\fi 
 \mathchardef#1="#3\next@#4#5} 
\def\mathhexbox@#1#2#3{\relax 
 \ifmmode\mathpalette{}{\m@th\mathchar"#1#2#3}%
 \else\leavevmode\hbox{$\m@th\mathchar"#1#2#3$}\fi} 

\def\bi#1{{\fam\bdifam\relax#1}} 
%
%
\ifams\input amsmacro\fi 
%
%
\newsymbol\bitGamma 3000 
\newsymbol\bitDelta 3001 
\newsymbol\bitTheta 3002 
\newsymbol\bitLambda 3003 
\newsymbol\bitXi 3004 
\newsymbol\bitPi 3005 
\newsymbol\bitSigma 3006 
\newsymbol\bitUpsilon 3007 
\newsymbol\bitPhi 3008 
\newsymbol\bitPsi 3009 
\newsymbol\bitOmega 300A 
\newsymbol\balpha 300B 
\newsymbol\bbeta 300C 
\newsymbol\bgamma 300D 
\newsymbol\bdelta 300E 
\newsymbol\bepsilon 300F 
\newsymbol\bzeta 3010 
\newsymbol\bfeta 3011 
\newsymbol\btheta 3012 
\newsymbol\biota 3013 
\newsymbol\bkappa 3014 
\newsymbol\blambda 3015 
\newsymbol\bmu 3016 
\newsymbol\bnu 3017 
\newsymbol\bxi 3018 
\newsymbol\bpi 3019 
\newsymbol\brho 301A 
\newsymbol\bsigma 301B 
\newsymbol\btau 301C 
\newsymbol\bupsilon 301D 
\newsymbol\bphi 301E 
\newsymbol\bchi 301F 
\newsymbol\bpsi 3020 
\newsymbol\bomega 3021 
\newsymbol\bvarepsilon 3022 
\newsymbol\bvartheta 3023 
\newsymbol\bvaromega 3024 
\newsymbol\bvarrho 3025 
\newsymbol\bvarzeta 3026 
\newsymbol\bvarphi 3027 
\newsymbol\bpartial 3040 
\newsymbol\bell 3060 
\newsymbol\bimath 307B 
\newsymbol\bjmath 307C 
\mathchardef\binfty "0\bsyfam@31 
\mathchardef\bnabla "0\bsyfam@72 
\mathchardef\bdot "2\bsyfam@01 
\mathchardef\bGamma "0\bffam@00 
\mathchardef\bDelta "0\bffam@01 
\mathchardef\bTheta "0\bffam@02 
\mathchardef\bLambda "0\bffam@03 
\mathchardef\bXi "0\bffam@04 
\mathchardef\bPi "0\bffam@05 
\mathchardef\bSigma "0\bffam@06 
\mathchardef\bUpsilon "0\bffam@07 
\mathchardef\bPhi "0\bffam@08 
\mathchardef\bPsi "0\bffam@09 
\mathchardef\bOmega "0\bffam@0A 
\mathchardef\itGamma "0100 
\mathchardef\itDelta "0101 
\mathchardef\itTheta "0102 
\mathchardef\itLambda "0103 
\mathchardef\itXi "0104 
\mathchardef\itPi "0105 
\mathchardef\itSigma "0106 
\mathchardef\itUpsilon "0107 
\mathchardef\itPhi "0108 
\mathchardef\itPsi "0109 
\mathchardef\itOmega "010A 
\mathchardef\Gamma "0000 
\mathchardef\Delta "0001 
\mathchardef\Theta "0002 
\mathchardef\Lambda "0003 
\mathchardef\Xi "0004 
\mathchardef\Pi "0005 
\mathchardef\Sigma "0006 
\mathchardef\Upsilon "0007 
\mathchardef\Phi "0008 
\mathchardef\Psi "0009 
\mathchardef\Omega "000A 
%
%
\newcount\firstpage  \firstpage=1  
\newcount\jnl                      
\newcount\secno                    
\newcount\subno                    
\newcount\subsubno                 
\newcount\appno                    
\newcount\tabno                    
\newcount\figno                    
\newcount\countno                  
\newcount\refno                    
\newcount\eqlett     \eqlett=97    
\newif\ifletter 
\newif\ifwide 
\newif\ifnotfull 
\newif\ifaligned 
\newif\ifnumbysec   
\newif\ifappendix 
\newif\ifnumapp 
\newif\ifssf 
\newif\ifppt 
\newdimen\t@bwidth 
\newdimen\c@pwidth 
\newdimen\digitwidth                    
\newdimen\argwidth                      
\newdimen\secindent    \secindent=5pc   
\newdimen\textind    \textind=16pt      
\newdimen\tempval                       
\newskip\beforesecskip 
\def\beforesecspace{\vskip\beforesecskip\relax} 
\newskip\beforesubskip 
\def\beforesubspace{\vskip\beforesubskip\relax} 
\newskip\beforesubsubskip 
\def\beforesubsubspace{\vskip\beforesubsubskip\relax} 
\newskip\secskip 
\def\secspace{\vskip\secskip\relax} 
\newskip\subskip 
\def\subspace{\vskip\subskip\relax} 
\newskip\insertskip 
\def\insertspace{\vskip\insertskip\relax} 
\def\sp@ce{\ifx\next*\let\next=\@ssf 
               \else\let\next=\@nossf\fi\next} 
\def\@ssf#1{\nobreak\secspace\global\ssftrue\nobreak} 
\def\@nossf{\nobreak\secspace\nobreak\noindent\ignorespaces} 
\def\subsp@ce{\ifx\next*\let\next=\@sssf 
               \else\let\next=\@nosssf\fi\next} 
\def\@sssf#1{\nobreak\subspace\global\ssftrue\nobreak} 
\def\@nosssf{\nobreak\subspace\nobreak\noindent\ignorespaces} 
\beforesecskip=24pt plus12pt minus8pt 
\beforesubskip=12pt plus6pt minus4pt 
\beforesubsubskip=12pt plus6pt minus4pt 
\secskip=12pt plus 2pt minus 2pt 
\subskip=6pt plus3pt minus2pt 
\insertskip=18pt plus6pt minus6pt%
\fontdimen16\tensy=2.7pt 
\fontdimen17\tensy=2.7pt 
%
%
\def\eqlabel{(\ifappendix\applett 
               \ifnumbysec\ifnum\secno>0 \the\secno\fi.\fi 
               \else\ifnumbysec\the\secno.\fi\fi\the\countno)} 
\def\seclabel{\ifappendix\ifnumapp\else\applett\fi 
    \ifnum\secno>0 \the\secno 
    \ifnumbysec\ifnum\subno>0.\the\subno\fi\fi\fi 
    \else\the\secno\fi\ifnum\subno>0.\the\subno 
         \ifnum\subsubno>0.\the\subsubno\fi\fi} 
\def\tablabel{\ifappendix\applett\fi\the\tabno} 
\def\figlabel{\ifappendix\applett\fi\the\figno} 
\def\gac{\global\advance\countno by 1} 
%
%
 
\def\vfootnote#1{\insert\footins\bgroup 
\interlinepenalty=\interfootnotelinepenalty 
\splittopskip=\ht\strutbox 
\splitmaxdepth=\dp\strutbox \floatingpenalty=20000 
\leftskip=0pt \rightskip=0pt \spaceskip=0pt \xspaceskip=0pt%
\noindent\smallfonts\rm #1\ \ignorespaces\footstrut\futurelet\next\fo@t} 
%
%
\def\endinsert{\egroup 
    \if@mid \dimen@=\ht0 \advance\dimen@ by\dp0 
       \advance\dimen@ by12\p@ \advance\dimen@ by\pagetotal 
       \ifdim\dimen@>\pagegoal \@midfalse\p@gefalse\fi\fi 
    \if@mid \insertspace \box0 \par \ifdim\lastskip<\insertskip 
    \removelastskip \penalty-200 \insertspace \fi 
    \else\insert\topins{\penalty100 
       \splittopskip=0pt \splitmaxdepth=\maxdimen  
       \floatingpenalty=0 
       \ifp@ge \dimen@=\dp0 
       \vbox to\vsize{\unvbox0 \kern-\dimen@}%
       \else\box0\nobreak\insertspace\fi}\fi\endgroup}    
%
%
%
\def\ind{\hbox to \secindent{\hfill}} 
%
%

%
%
 
%
%
\def\indeqn#1{\alignedfalse\displ@y\halign{\hbox to \displaywidth 
    {$\ind\@lign\displaystyle##\hfil$}\crcr #1\crcr}} 
%
%
\def\indalign#1{\alignedtrue\displ@y \tabskip=0pt  
  \halign to\displaywidth{\ind$\@lign\displaystyle{##}$\tabskip=0pt 
    &$\@lign\displaystyle{{}##}$\hfill\tabskip=\centering 
    &\llap{$\@lign\hbox{\rm##}$}\tabskip=0pt\crcr 
    #1\crcr}} 
\def\fl{{\hskip-\secindent}} 
\def\indenteddisplay#1$${\indispl@y{#1 }} 
\def\indispl@y#1{\disptest#1\eqalignno\eqalignno\disptest} 
\def\disptest#1\eqalignno#2\eqalignno#3\disptest{%
    \ifx#3\eqalignno 
    \indalign#2%
    \else\indeqn{#1}\fi$$} 
%
%
 
%
%
 
%
%
 
%
%
 
%
%

\def\ns{\noalign{\vskip-3pt}}

%
 
%
%
\def\bhbar{\rlap{\kern1pt\raise.4ex\hbox{\bf\char'40}}\bi{h}} 
\def\case#1#2{{\textstyle{#1\over#2}}}

\def\e{{\rm e}} 
 
\def\frac#1#2{{#1\over#2}} 
\ifams 
\def\lap{\lesssim} 
\def\gap{\gtrsim} 
\let\le=\leqslant 
 
\let\leq=\leqslant 
\let\ge=\geqslant

\else

\def\gap{\;\lower3pt\hbox{$\buildrel > \over \sim$}\;}%
\def\lap{\;\lower3pt\hbox{$\buildrel < \over \sim$}\;}\fi 
 
\chardef\ii="10 
\def\tqs{\hbox to 25pt{\hfil}}

\def\Bbbone{1\kern-.22em {\rm l}} 
%
%
\def\rp{\raise8pt\hbox{$\scriptstyle\prime$}} 
%
%
%
%

%
%
\def\[#1\]{\setbox0=\hbox{$\dsty#1$}\argwidth=\wd0 
    \setbox0=\hbox{$\left[\box0\right]$}\advance\argwidth by -\wd0 
    \left[\kern.3\argwidth\box0\kern.3\argwidth\right]} 
%
%
\def\lsb#1\rsb{\setbox0=\hbox{$#1$}\argwidth=\wd0 
    \setbox0=\hbox{$\left[\box0\right]$}\advance\argwidth by -\wd0 
    \left[\kern.3\argwidth\box0\kern.3\argwidth\right]} 
%
 
%
%
 
%
\def\pt(#1){({\it #1\/})} 
\let\dsty=\displaystyle

%
%
\def\reactions#1{\vskip 12pt plus2pt minus2pt%
\vbox{\hbox{\kern\secindent\vrule\kern12pt%
\vbox{\kern0.5pt\vbox{\hsize=24pc\parindent=0pt\smallfonts\rm NUCLEAR  
REACTIONS\strut\quad #1\strut}\kern0.5pt}\kern12pt\vrule}}} 
%
%
\def\slashchar#1{\setbox0=\hbox{$#1$}\dimen0=\wd0%
\setbox1=\hbox{/}\dimen1=\wd1%
\ifdim\dimen0>\dimen1%
\rlap{\hbox to \dimen0{\hfil/\hfil}}#1\else                                         
\rlap{\hbox to \dimen1{\hfil$#1$\hfil}}/\fi} 
%
%
\def\textindent#1{\noindent\hbox to \parindent{#1\hss}\ignorespaces} 
%
%
\def\opencirc{\raise1pt\hbox{$\scriptstyle{\bigcirc}$}} 
 
\ifams 
\def\opensqr{\hbox{$\square$}} 
 
\def\opentridown{\hbox{$\triangledown$}}

\else 
\def\opensqr{\vbox{\hrule height.4pt\hbox{\vrule width.4pt height3.5pt 
    \kern3.5pt\vrule width.4pt}\hrule height.4pt}} 
 
\def\opentridown{\raise1pt\hbox{$\scriptstyle\bigtriangledown$}}

\fi

%
%
\def\m@th{\mathsurround=0pt} 
%
%
\def\cases#1{%
\left\{\,\vcenter{\normalbaselines\openup1\jot\m@th%
     \ialign{$\displaystyle##\hfil$&\rm\tqs##\hfil\crcr#1\crcr}}\right.}%
%
%
\def\oldcases#1{\left\{\,\vcenter{\normalbaselines\m@th 
    \ialign{$##\hfil$&\rm\quad##\hfil\crcr#1\crcr}}\right.} 
%
%
\def\numcases#1{\left\{\,\vcenter{\baselineskip=15pt\m@th%
     \ialign{$\displaystyle##\hfil$&\rm\tqs##\hfil 
     \crcr#1\crcr}}\right.\hfill 
     \vcenter{\baselineskip=15pt\m@th%
     \ialign{\rlap{$\phantom{\displaystyle##\hfil}$}\tabskip=0pt&\en 
     \rlap{\phantom{##\hfil}}\crcr#1\crcr}}} 
\def\ptnumcases#1{\left\{\,\vcenter{\baselineskip=15pt\m@th%
     \ialign{$\displaystyle##\hfil$&\rm\tqs##\hfil 
     \crcr#1\crcr}}\right.\hfill 
     \vcenter{\baselineskip=15pt\m@th%
     \ialign{\rlap{$\phantom{\displaystyle##\hfil}$}\tabskip=0pt&\enpt 
     \rlap{\phantom{##\hfil}}\crcr#1\crcr}}\global\eqlett=97 
     \global\advance\countno by 1} 
%
%
\def\eq(#1){\ifaligned\@mp(#1)\else\hfill\llap{{\rm (#1)}}\fi} 
\def\ceq(#1){\ns\ns\ifaligned\@mp\fi\eq(#1)\cr\ns\ns} 
\def\eqpt(#1#2){\ifaligned\@mp(#1{\it #2\/}) 
                    \else\hfill\llap{{\rm (#1{\it #2\/})}}\fi} 
 
%
%
\countno=1 
\def\eqnobysec{\numbysectrue} 
\def\aleq{&\rm(\ifappendix\applett 
               \ifnumbysec\ifnum\secno>0 \the\secno\fi.\fi 
               \else\ifnumbysec\the\secno.\fi\fi\the\countno} 
\def\noaleq{\hfill\llap\bgroup\rm(\ifappendix\applett 
               \ifnumbysec\ifnum\secno>0 \the\secno\fi.\fi 
               \else\ifnumbysec\the\secno.\fi\fi\the\countno} 
\def\@mp{&} 
\def\en{\ifaligned\aleq)\else\noaleq)\egroup\fi\gac} 
\def\cen{\ns\ns\ifaligned\@mp\fi\en\cr\ns\ns} 
\def\enpt{\ifaligned\aleq{\it\char\the\eqlett})\else 
    \noaleq{\it\char\the\eqlett})\egroup\fi 
    \global\advance\eqlett by 1} 
\def\endpt{\ifaligned\aleq{\it\char\the\eqlett})\else 
    \noaleq{\it\char\the\eqlett})\egroup\fi 
    \global\eqlett=97\gac} 
%
%

\def\JPA{{\it J. Phys. A: Math. Gen.}} 
\def\JPC{{\it J. Phys. C: Solid State Phys.}}     

 

%
%

\def\APNY{{\it Ann. Phys., NY\/}}

\def\PR{{\it Phys. Rev.}} 
\def\PRL{{\it Phys. Rev. Lett.}}

\def\RMP{{\it Rev. Mod. Phys.}}

\headline={\ifodd\pageno{\ifnum\pageno=\firstpage\hfill 
   \else\rrhead\fi}\else\lrhead\fi} 
\def\rrhead{\textfonts\hskip\secindent\it 
    \shorttitle\hfill\rm\folio} 
\def\lrhead{\textfonts\hbox to\secindent{\rm\folio\hss}%
    \it\aunames\hss} 
\footline={\ifnum\pageno=\firstpage \hfill\textfonts\rm\folio\fi} 
\def\@rticle#1#2{\vglue.5pc 
    {\parindent=\secindent \bf #1\par} 
     \vskip2.5pc 
    {\exhyphenpenalty=10000\hyphenpenalty=10000 
     \baselineskip=18pt\raggedright\noindent 
     \headfonts\bf#2\par}\futurelet\next\sh@rttitle}%
\def\title#1{\gdef\shorttitle{#1} 
    \vglue4pc{\exhyphenpenalty=10000\hyphenpenalty=10000  
    \baselineskip=18pt  
    \raggedright\parindent=0pt 
    \headfonts\bf#1\par}\futurelet\next\sh@rttitle}  

\def\article#1#2{\gdef\shorttitle{#2}\@rticle{#1}{#2}}  
\def\review#1{\gdef\shorttitle{#1}%
    \@rticle{REVIEW \ifpbm\else ARTICLE\fi}{#1}} 
\def\topical#1{\gdef\shorttitle{#1}%
    \@rticle{TOPICAL REVIEW}{#1}} 
\def\comment#1{\gdef\shorttitle{#1}%
    \@rticle{COMMENT}{#1}} 
\def\note#1{\gdef\shorttitle{#1}%
    \@rticle{NOTE}{#1}} 
\def\prelim#1{\gdef\shorttitle{#1}%
    \@rticle{PRELIMINARY COMMUNICATION}{#1}} 
\def\letter#1{\gdef\shorttitle{Letter to the Editor}%
     \gdef\aunames{Letter to the Editor} 
     \global\lettertrue\ifnum\jnl=7\global\letterfalse\fi 
     \@rticle{LETTER TO THE EDITOR}{#1}} 
\def\sh@rttitle{\ifx\next[\let\next=\sh@rt 
                \else\let\next=\f@ll\fi\next} 
\def\sh@rt[#1]{\gdef\shorttitle{#1}} 
\def\f@ll{} 
\def\author#1{\ifletter\else\gdef\aunames{#1}\fi\vskip1.5pc 
    {\parindent=\secindent   
     \hang\textfonts   
     \ifppt\bf\else\rm\fi#1\par}   
     \ifppt\bigskip\else\smallskip\fi 
     \futurelet\next\@unames} 
\def\@unames{\ifx\next[\let\next=\short@uthor 
                 \else\let\next=\@uthor\fi\next} 
\def\short@uthor[#1]{\gdef\aunames{#1}} 
\def\@uthor{} 
\def\address#1{{\parindent=\secindent 
    \exhyphenpenalty=10000\hyphenpenalty=10000 
\ifppt\textfonts\else\smallfonts\fi\hang\raggedright\rm#1\par}%
    \ifppt\bigskip\fi} 
\def\jl#1{\global\jnl=#1} 
\jl{0}%
\def\journal{\ifnum\jnl=1 J. Phys.\ A: Math.\ Gen.\  
        \else\ifnum\jnl=2 J. Phys.\ B: At.\ Mol.\ Opt.\ Phys.\  
        \else\ifnum\jnl=3 J. Phys.:\ Condens. Matter\  
        \else\ifnum\jnl=4 J. Phys.\ G: Nucl.\ Part.\ Phys.\  
        \else\ifnum\jnl=5 Inverse Problems\  
        \else\ifnum\jnl=6 Class. Quantum Grav.\  
        \else\ifnum\jnl=7 Network\  
        \else\ifnum\jnl=8 Nonlinearity\ 
        \else\ifnum\jnl=9 Quantum Opt.\ 
        \else\ifnum\jnl=10 Waves in Random Media\ 
        \else\ifnum\jnl=11 Pure Appl. Opt.\  
        \else\ifnum\jnl=12 Phys. Med. Biol.\ 
        \else\ifnum\jnl=13 Modelling Simulation Mater.\ Sci.\ Eng.\  
        \else\ifnum\jnl=14 Plasma Phys. Control. Fusion\  
        \else\ifnum\jnl=15 Physiol. Meas.\  
        \else\ifnum\jnl=16 Sov.\ Lightwave Commun.\ 
        \else\ifnum\jnl=17 J. Phys.\ D: Appl.\ Phys.\ 
        \else\ifnum\jnl=18 Supercond.\ Sci.\ Technol.\ 
        \else\ifnum\jnl=19 Semicond.\ Sci.\ Technol.\ 
        \else\ifnum\jnl=20 Nanotechnology\ 
        \else\ifnum\jnl=21 Meas.\ Sci.\ Technol.\  
        \else\ifnum\jnl=22 Plasma Sources Sci.\ Technol.\  
        \else\ifnum\jnl=23 Smart Mater.\ Struct.\  
        \else\ifnum\jnl=24 J.\ Micromech.\ Microeng.\ 
   \else Institute of Physics Publishing\  
   \fi\fi\fi\fi\fi\fi\fi\fi\fi\fi\fi\fi\fi\fi\fi 
   \fi\fi\fi\fi\fi\fi\fi\fi\fi} 
\let\abs=\beginabstract 

\let\endabs=\endabstract 
\def\submitted{\ifppt\noindent\textfonts\rm Submitted to \journal\par 
     \bigskip\fi} 
\def\today{\number\day\ \ifcase\month\or 
     January\or February\or March\or April\or May\or June\or 
     July\or August\or September\or October\or November\or 
     December\fi\space \number\year} 
\def\date{\ifppt\noindent\textfonts\rm  
     Date: \today\par\goodbreak\bigskip\fi} 
%
%
\def\pacs#1{\ifppt\noindent\textfonts\rm  
     PACS number(s): #1\par\bigskip\fi} 
%
 
%
%
\def\section#1{\ifppt\ifnum\secno=0\eject\fi\fi 
    \subno=0\subsubno=0\global\advance\secno by 1 
    \gdef\labeltype{\seclabel}\ifnumbysec\countno=1\fi 
    \goodbreak\beforesecspace\nobreak 
    \noindent{\bf \the\secno. #1}\par\futurelet\next\sp@ce} 
\def\subsection#1{\subsubno=0\global\advance\subno by 1 
     \gdef\labeltype{\seclabel}%
     \ifssf\else\goodbreak\beforesubspace\fi 
     \global\ssffalse\nobreak 
     \noindent{\it \the\secno.\the\subno. #1\par}%
     \futurelet\next\subsp@ce} 
\def\subsubsection#1{\global\advance\subsubno by 1 
     \gdef\labeltype{\seclabel}%
     \ifssf\else\goodbreak\beforesubsubspace\fi 
     \global\ssffalse\nobreak 
     \noindent{\it \the\secno.\the\subno.\the\subsubno. #1}\null.  
     \ignorespaces} 
%
 
%
%
\def\numappendix#1{\ifappendix\ifnumbysec\countno=1\fi\else 
    \countno=1\figno=0\tabno=0\fi 
    \subno=0\global\advance\appno by 1 
    \secno=\appno\gdef\applett{A}\gdef\labeltype{\seclabel}%
    \global\appendixtrue\global\numapptrue 
    \goodbreak\beforesecspace\nobreak 
    \noindent{\bf Appendix \the\appno. #1\par}%
    \futurelet\next\sp@ce} 
\def\numsubappendix#1{\global\advance\subno by 1\subsubno=0 
    \gdef\labeltype{\seclabel}%
    \ifssf\else\goodbreak\beforesubspace\fi 
    \global\ssffalse\nobreak 
    \noindent{\it A\the\appno.\the\subno. #1\par}%
    \futurelet\next\subsp@ce} 
\def\@ppendix#1#2#3{\countno=1\subno=0\subsubno=0\secno=0\figno=0\tabno=0 
    \gdef\applett{#1}\gdef\labeltype{\seclabel}\global\appendixtrue 
    \goodbreak\beforesecspace\nobreak 
    \noindent{\bf Appendix#2#3\par}\futurelet\next\sp@ce} 
\def\Appendix#1{\@ppendix{A}{. }{#1}} 
\def\appendix#1#2{\@ppendix{#1}{ #1. }{#2}} 
\def\App#1{\@ppendix{A}{ }{#1}} 
\def\app{\@ppendix{A}{}{}} 
\def\subappendix#1#2{\global\advance\subno by 1\subsubno=0 
    \gdef\labeltype{\seclabel}%
    \ifssf\else\goodbreak\beforesubspace\fi 
    \global\ssffalse\nobreak 
    \noindent{\it #1\the\subno. #2\par}%
    \nobreak\subspace\noindent\ignorespaces} 
%
%
\def\@ck#1{\ifletter\bigskip\noindent\ignorespaces\else 
    \goodbreak\beforesecspace\nobreak 
    \noindent{\bf Acknowledgment#1\par}%
    \nobreak\secspace\noindent\ignorespaces\fi} 
\def\ack{\@ck{s}} 
\def\ackn{\@ck{}} 
\def\n@ip#1{\goodbreak\beforesecspace\nobreak 
    \noindent\smallfonts{\it #1}. \rm\ignorespaces} 
\def\naip{\n@ip{Note added in proof}} 
\def\na{\n@ip{Note added}} 
 
%
%
 
%
 
%
%
 
%
 
%
 
\def\tablecont{\topinsert\global\advance\tabno by -1 
    \tablecaption{(continued)}} 
\def\tablecaption#1{\gdef\labeltype{\tablabel}\global\widefalse 
    \leftskip=\secindent\parindent=0pt 
    \global\advance\tabno by 1 
    \smallfonts{\bf Table \ifappendix\applett\fi\the\tabno.} \rm #1\par 
    \smallskip\futurelet\next\t@b} 
\def\t@b{\ifx\next*\let\next=\widet@b 
             \else\ifx\next[\let\next=\fullwidet@b 
                      \else\let\next=\narrowt@b\fi\fi 
             \next} 
\def\widet@b#1{\global\widetrue\global\notfulltrue 
    \t@bwidth=\hsize\advance\t@bwidth by -\secindent}  
\def\fullwidet@b[#1]{\global\widetrue\global\notfullfalse 
    \leftskip=0pt\t@bwidth=\hsize}                   
\def\narrowt@b{\global\notfulltrue} 
\def\align{\catcode`?=13\ifnotfull\moveright\secindent\fi 
    \vbox\bgroup\halign\ifwide to \t@bwidth\fi 
    \bgroup\strut\tabskip=1.2pc plus1pc minus.5pc} 
\def\endalign{\egroup\egroup\catcode`?=12} 
 
%
%

%
%

%
 
%
%

%
 
\catcode`?=13 
\def\lineup{\setbox0=\hbox{\smallfonts\rm 0}%
    \digitwidth=\wd0%
    \def?{\kern\digitwidth}%
    \def\\{\hbox{$\phantom{-}$}}%
    \def\-{\llap{$-$}}} 
\catcode`?=12 
%
%
\def\sidetable#1#2{\hbox{\ifppt\hsize=18pc\t@bwidth=18pc 
                          \else\hsize=15pc\t@bwidth=15pc\fi 
    \parindent=0pt\vtop{\null #1\par}%
    \ifppt\hskip1.2pc\else\hskip1pc\fi 
    \vtop{\null #2\par}}}  
\def\lstable#1#2{\everypar{}\tempval=\hsize\hsize=\vsize 
    \vsize=\tempval\hoffset=-3pc 
    \global\tabno=#1\gdef\labeltype{\tablabel}%
    \noindent\smallfonts{\bf Table \ifappendix\applett\fi 
    \the\tabno.} \rm #2\par 
    \smallskip\futurelet\next\t@b} 
\def\inctabno{\global\advance\tabno by 1} 
%
%
 
%
 
%
\def\figure#1{\figc@ption{#1}\bigskip} 
\def\figc@ption#1{\global\advance\figno by 1\gdef\labeltype{\figlabel}%
   {\parindent=\secindent\smallfonts\hang 
    {\bf Figure \ifappendix\applett\fi\the\figno.} \rm #1\par}} 
%
%
\def\refHEAD{\goodbreak\beforesecspace 
     \noindent\textfonts{\bf References}\par 
     \let\ref=\rf 
     \nobreak\smallfonts\rm} 
\def\numreferences{\refHEAD\parindent=30pt 
     \everypar{\hang\noindent\frenchspacing\rm} 
     \secspace} 
\def\rf#1{\par\noindent\hbox to 21pt{\hss #1\quad}\ignorespaces} 
%
 
%
 
%
%
 
%
%
 
%
%

%
%

%
\catcode`\@=12 
%
%
 
%
%
\def\jnlstyle{\pptfalse\headsize{14}{18}%
\textsize{10}{12}%
\smallsize{8}{10} 
\textind=16pt} 
%
%
 
%
%
 
%
\parindent=\textind 
\input xref
\input epsf
\eqnobysec
\def\figure#1{\global\advance\figno by 1\gdef\labeltype{\figlabel}%
   {\parindent=\secindent\smallfonts\hang 
    {\bf Figure \ifappendix\applett\fi\the\figno.} \rm #1\par}} 
 


\jnlstyle

\jl{1}
    
\overfullrule=0pt

\title{Extended surface disorder in the
quantum Ising chain}[Extended surface disorder in the quantum Ising
chain]

\author{L Turban\dag, D Karevski\dag\ and F Igl\'oi\ddag}[L Turban, D
Karevski and F Igl\'oi]

\address{\dag Laboratoire de Physique des 
Mat\'eriaux\footnote{\S}{Unit\'e Mixte de 
Recherche CNRS no 7556}, Universit\'e Henri Poincar\'e (Nancy~I),
BP~239,  F--54506~Vand\oe uvre l\`es Nancy Cedex, France}

\address{\ddag Research Institute for Solid State Physics and
Optics, PO Box 49, H--1525 Budapest, Hungary and Institute of
Theoretical Physics, Szeged University, H--6720 Szeged,
Hungary}


\abs
We consider random extended surface perturbations in the
transverse field Ising model decaying as a power of the
distance from the surface towards a pure bulk system.
The decay may be linked either to the evolution of the couplings or to
their probabilities. Using scaling arguments, we develop a
relevance-irrelevance criterion for such perturbations. We study the
probability distribution of the surface magnetization, its average and
typical critical behaviour for marginal and relevant perturbations.
According to analytical results, the surface magnetization follows a
log-normal distribution and both the average and typical critical
behaviours are characterized by power-law singularities with
continuously varying exponents in the marginal case and essential
singularities in the relevant case. For enhanced average local
couplings, the transition becomes first order with a nonvanishing
critical surface magnetization. This occurs above a positive threshold
value of the perturbation amplitude in the marginal case. 
\endabs 


\pacs{05.50.+q, 64.60.Fr, 68.35.Rh}

\submitted

\date
\section{Introduction}
Quenched, i.e., time-independent disorder has a strong influence
on the nature of quantum phase transitions which take place at zero
temperature \cite{rieger97a}. Many interesting features of the effect of
randomness can be observed in one-dimensional ($1d$) systems for which
several exact results have been obtained. Recently, the $1d$ random
transverse-field Ising model (TIM) has been the subject of much
interest. It is defined by the Hamiltonian:  
$$ 
{\cal
H}=-{1\over2}\sum_l(J_l\,\sigma^z_l\sigma^z_{l+1}+h_l\sigma^x_l)\,,
\label{e1.1}  
$$
where $\sigma^x_l$ and $\sigma^z_l$ are Pauli matrices at site $l$
and the exchange couplings, $J_l$, the transverse fields, $h_l$,
are random variables. 

For homogeneous independently distributed couplings and
fields, Fisher \cite{fisher92,fisher95} obtained many new results about
the static properties of the random TIM in~\ref{e1.1} using a real space
renormalization group method which is believed to become exact
at large scales, i.e., sufficiently close to the critical point.
Later, Fisher's conjectures have been checked 
numerically [4--6] and new results have been
obtained, especially about the dynamical properties of the model at the
critical point \cite{rieger97b,igloi98a} as well as in the Griffiths 
phase
\cite{igloi98b,igloi99a}.

In many physical situation the disorder is not homogeneous. A free
surface or a defect in the bulk can evidently induce a disorder
which is inhomogeneously distributed in its vicinity. In the
present paper we shall consider a random version of the
Hilhorst-van Leeuwen (HvL) model [11--13]. In
this model, the couplings are modified near a free surface and the
amplitude of the perturbation decays according to a power law
$l^{-\kappa}$ where $l$ denotes the distance from the
surface. A relevance-irrelevance criterion has been proposed for
such smoothly inhomogeneous perturbations, showing that marginal
behaviour is obtained when the decay exponent $\kappa=1/\nu$
where $\nu$ is the correlation length exponent of the pure
system [14--16]. The critical behaviour is then characterized by local
exponents which vary continuously with the perturbation amplitude. For a
slower decay the perturbation becomes relevant and, locally, power laws
are replaced by essential singularities.   

We consider two types of random extended surface
perturbations associated with a pure bulk system: 

For the first type of perturbation, the probabilities $p_l$ and $q_l$
are constant and the couplings take the values:
$$
J_l=J\left[1+a_1{(-1)^{f_l}\over l^\kappa}\right]\,,\quad
f_l=\left\{
\matrix{
1&\quad p_l=1/2\cr
0&\quad q_l=1/2\cr
}\right.
\qquad{\rm (model\ \#1)} 
\label{e1.2} 
$$
The perturbation has a vanishing average and the couplings decay
towards their constant bulk value $J$.

For the second type, the decay is associated with the
probabilities whereas the amplitudes remain constant. The couplings
follow the binary distribution:
$$
J_l=J(1+f_l a_2)\,,\qquad
f_l=\left\{
\matrix{
1&\quad p_l=l^{-\kappa}\cr
0&\quad q_l=1-l^{-\kappa}\cr
}\right.
\qquad{\rm (model\ \#2)} 
\label{e1.3} 
$$
The average coupling takes the same form as in the HvL
model and one recovers asymptotically a pure bulk system with couplings
equal to $J$.

The transverse field is always assumed to be constant, $h_l=h=1$, thus 
the
bulk is critical at $J=1$.  

In the present work, we study the surface magnetic properties in
the marginal situation which occurs at
$\kappa=\case{1}{2}$ for model \#1 and $\kappa=1$ for model \#2 as well
as for relevant perturbations corresponding to slower decays. 

The same type of problem with an aperiodic distribution of the couplings
following some substitution sequence has been considered recently
\cite{turban99}.

The structure of the paper is the following. The relevance of random
extended surface perturbations is discussed in section~2. In section~3,
analytical results about the surface magnetization of the inhomogeneous
TIM are presented.
The two models of inhomogeneous surface disorder are studied in 
section~4 for marginal perturbations and section~5 for relevant
ones. The results are discussed in section~6.

\section{Relevance-irrelevance criterion}

We begin with a discussion of the relevance of random extended surface
perturbations in a $d$-dimensional classical system. The unperturbed
system has a free surface of dimension $d-1$ at $l=0$, perpendicular to
the unit vector $\bi{u}$, and the position vector is written as
$\bi{r}=l\bi{u}+\bi{r}_\parallel$. The layered perturbation, which
couples to the operator density $\phi(\bi{r})$ with bulk scaling
dimension $x_{\phi}$, is written as:
$$
-\beta V=\sum_\bi{r}A(l)\phi(\bi{r})\,,
\label{e2.1}
$$
where $\beta=(k_{\rm B}T)^{-1}$. The perturbation amplitude
$A(l)$ in the different layers are independent random variables
such that:  
$$
\fl\eqalign{
&[A(l)]_{\rm av}=0\,,\qquad[A(l)A(l')]_{\rm av}
={A_1^2\over l^{2\kappa}}\,\delta_{ll'}\qquad{\rm (model\ \#1)}\cr
&[A(l)]_{\rm av}={A_2\over l^\kappa}\,,\qquad[A(l)A(l')]_{\rm av}
={A_2^2\over l^\kappa}\,\delta_{ll'}+{A_2^2\over(ll')^\kappa}\,(1-\delta_{ll'})\qquad{\rm (model\ \#2)}\cr
}
\label{e2.2}
$$
Thermodynamic perturbation theory, up to second order, gives the
following correction to the free energy: 
$$
\eqalign{
-\beta\Delta F&=-\beta\langle V\rangle
+{1\over2}\beta^2\left[\langle V^2\rangle-\langle V\rangle^2\right]\cr 
              &=\sum_\bi{r}A(l)\langle\phi(\bi{r})\rangle
+{1\over2}\sum_{\bi{r},\bi{r}'}A(l)A(l'){\cal
G}_{\phi\phi}(\bi{r},\bi{r}')\,,\cr
}
\label{e2.3}
$$
where $\langle\cdots\rangle$ denotes a thermal average and ${\cal
G}_{\phi\phi}$ is the connected two-point correlation function of the
operator $\phi(\bi{r})$, both for the unperturbed system.

In the case of model \#1, the average of the first-order correction
vanishes. The relevance of the perturbation is linked to the
scaling behaviour of the amplitude in the average of the second-order
correction: 
$$
-\beta[\Delta
F^{(2)}]_{\rm av}={L^{d-1}\over2}\sum_l\sum_{\bi{r}_\parallel}
{A_1^2\over l^{2\kappa}}\,{\cal G}_{\phi\phi}(r_\parallel)\,,
\label{e2.4} $$
where $L^{d-1}$ is the surface of the layers.
The average perturbation density has dimension $2d-1$ wheras the
correlation function has dimension $2x_\phi$, thus, under a length
rescaling by a factor $b=L/L'$,
$$
{{A'}_1^2\over l'^{2\kappa}}=b^{2d-1-2x_\phi}{A_1^2\over l^{2\kappa}}
=b^{2y_\phi-1}{A_1^2\over l^{2\kappa}}
\label{e2.5}
$$
and the second order perturbation amplitude transforms as:
$$
{A'}_1^2=b^{2y_\phi-2\kappa-1}A_1^2\,,
\label{e2.6}
$$
where $y_\phi=d-x_\phi$ is the bulk scaling dimension of the variable
conjugate to $\phi$. It follows that the perturbation is marginal when
$\kappa=\kappa_1=y_\phi-\case{1}{2}$. It becomes relevant (irrelevant)
when $\kappa<(>)\kappa_1$.

In the case of model \#2, the average of the first-order correction is given
by:  
$$
-\beta[\Delta F^{(1)}]_{\rm av}
=\sum_\bi{r}{A_2\over l^\kappa}\langle\phi(\bi{r})\rangle\,.
\label{e2.7}
$$
The average perturbation density has now dimension $d$ so that
$$ 
{{A'}_2\over l'^\kappa}=b^{d-x_\phi}{A_2\over l^\kappa}
=b^{y_\phi}{A_2\over l^\kappa}
\label{e2.8}
$$
and:
$$
{A'}_2=b^{y_\phi-\kappa}A_2\,.
\label{e2.9}
$$
The perturbation is marginal when $\kappa=\kappa_2=y_\phi$,
relevant (irrelevant) when $\kappa<(>)\kappa_2$. One may notice that the
average perturbation is of the HvL form, thus the relevance criterion is the
same for both models to first order in the perturbation amplitude. A scaling
analysis of the second-order correction contributed by the off-diagonal
term in $[A(l)A(l')]_{\rm av}$ leads to the same conclusion. 
Comparing the form of $[A(l)A(l')]_{\rm av}$ for both models, one concludes that
the diagonal second-order correction in model \#2 scales as $A_1^2$ in
equation~\ref{e2.6} with $2\kappa$ replaced by $\kappa$, i.e., as
$$
[{A'}_2^2]_{\rm diag}=b^{2y_\phi-\kappa-1}[A_2^2]_{\rm diag}\,.
\label{e2.10}
$$
It follows that the second-order diagonal correction is relevant when
$\kappa<2y_\phi-1$. It is more dangerous than the first-order correction when
$y_\phi>1$.  

In the case of the $2d$ Ising model with a thermal perturbation, such 
that $\phi(\bi{r})$ is the energy density $\sigma_\bi{r}\sigma_\bi{r+u}$, one
has $y_\phi=1/\nu=1$ and marginal behaviour is obtained for
$\kappa_1=\case{1}{2}$ and $\kappa_2=1$ as indicated in the
introduction. 

The $1d$ TIM, defined in equation~\ref{e1.1} together with \ref{e1.2}
or~\ref{e1.3}, which corresponds to the extreme anisotropic limit of the
$2d$ classical systems discussed here \cite{kogut79}, belongs to the 
same universality class, i.e., the relevance of the perturbation is the same
for both problems. 

\section{Surface magnetization and surface correlations}
Let us consider the imaginary time spin-spin autocorrelation
function of the TIM:
$$
{\cal G}_l(\tau)=\langle0\vert\sigma^z_l(\tau)
\sigma^z_l(0)\vert0\rangle
=\sum_i\vert\langle i\vert\sigma^z_l\vert0\rangle\vert^2
\exp[-\tau(E_i-E_0)]\,, 
\label{e3.1} 
$$
where $\vert0\rangle$ and $\vert i\rangle$ denotes the ground
state and the $i$th excited state of ${\cal H}$ in equation
\ref{e1.1}, with energies $E_0$ and $E_i$, respectively. With
symmetry breaking boundary conditions, i.e., fixing the spin at
the other end of the chain, such that $\sigma^z_L=\pm1$, the
ground state of the system is degenerate and in the large-$\tau$
limit we have $\lim_{\tau\to\infty}{\cal G}_l(\tau)=m_l^2$,
with a local magnetization given by:
$$
m_l=\langle1\vert\sigma^z_l\vert0\rangle\,.
\label{e3.2}
$$
To calculate $m_l$ we use the standard
methods of Lieb, Schultz and Mattis \cite{lieb61} and transform
${\cal H}$ into a free-fermion Hamiltonian,
$$
{\cal
H}=\sum_{q=1}^L\epsilon_q\left(\eta^\dagger_q\eta_q-{1\over2}\right)\,,
\label{e3.3} $$
in terms of the fermion creation (annihilation) operators
$\eta^\dagger_q$ ($\eta_q$). On a chain with length $L$ and free
boundary conditions, the non-negative fermion excitations
$\epsilon_q$ satisfy the set of equations
$$
\eqalign{&\epsilon_q\psi_q(l)=-h_l\varphi_q(l)-J_l\varphi_q(l+1)\,,\cr
&\epsilon_q\varphi_q(l)=-J_{l-1}\psi_q(l-1)-h_l\psi_q(l)\,,\cr}
\label{e3.4} 
$$
with $J_0=J_L=0$. Introducing the $2L$-dimensional vector
$\bi{V}_q$ with components
$$
V_q(2l-1)=-\varphi_q(l)\,,\qquad V_q(2l)=\psi_q(l)\,,
\label{e3.5}
$$
the relations in equation \ref{e3.4} lead to an eigenvalue
problem for the tridiagonal matrix: 
$$
{\bss T}=\pmatrix{0  &h_1&      &       &      &   \cr
	                 h_1& 0 &  J_1 &       &      &   \cr
                     &J_1&   0  &  h_2  &      &   \cr
                     &   &\ddots&\ddots &\ddots&   \cr
                     &   &      &J_{L-1}&\ddots&h_L\cr
                     &   &      &       & h_L  & 0 \cr}\,.
\label{e3.6}
$$
Taking the square of ${\bss T}$, odd and even components of
$\bi{V}_q$ decouple and one recovers two separate eigenvalue
problems for $\bvarphi$ and $\bpsi$. According to
equation~\ref{e3.4}, changing $\bvarphi_q$ into $-\bvarphi_q$ in
$\bi{V}_q$, the eigenvector corresponding to $-\epsilon_q$ is
obtained. Thus all the needed information is contained in that
part of the spectrum of ${\bss T}$ with $\epsilon_q\ge0$. Later on
we shall restrict ourselves to this sector.

Fixing the surface spin at $l=L$ amounts to put $h_L=0$ in $\bss
T$, thus the lowest excitation $\epsilon_1=0$ and the ground state
of the system is degenerate.

Using Wick's theorem, one may show that the local magnetization $m_l$,
with a fixed spin at $l=L$, is given by a determinant \cite{igloi98a}. 
For
the surface spin, at $l=1$, the expression simplifies considerably and
the surface magnetization takes the form \cite{peschel84}: 
$$
m_{\rm s}(L)=m_1=\varphi_1(1)=\left[1+\sum_{l=1}^{L-1}
\prod_{k=1}^l\left({h_k\over J_k}\right)^2\right]^{-{1\over2}}\,.
\label{e3.7}
$$
This can be rewritten under the equivalent form
$$
m_{\rm s}(L)=\overline{m}^{\rm d}_{\rm 
s}(L)\prod_{l=1}^{L-1}\left({J_l\over
h_l}\right)
\label{e3.8}
$$
where $\overline{m}^{\rm d}_{\rm s}$ is the surface magnetization at
$l=L$ for the dual chain (for which the fields and the couplings are
exchanged, $h_l\leftrightarrow J_l$) with a fixed spin now at $l=1$ 

We shall illustrate the usefulness of the last expression on the
example of the HvL model which corresponds to the Hamiltonian~\ref{e1.1}
with $h_l=1$ and $J_l=J(1+a l^{-\kappa})$. In the marginal
situation, $\kappa=1$, the product in~\ref{e3.8} asymptotically
behaves as $\prod_{l=1}^LJ_l\sim L^a$ at the critical point, $J_{\rm
c}=1$. Furthermore, when $L\gg1$, the couplings at the other end of the
chain are asymptotically unperturbed, thus one expects that
$\overline{m}^{\rm d}_{\rm s}(L)\sim L^{-{1\over2}}$, as for the
homogeneous model at the ordinary transition. Then from~\ref{e3.8} we
deduce the continuously varying dimension  
$$
x^{\rm s}_{\rm m}={{1\over2}-a}\,,\qquad a\le{1\over2}\,,
\label{e3.9}
$$
which governs the scaling behaviour of the surface magnetization, 
$m_{\rm
s}(L)\sim L^{-x^{\rm s}_{\rm m}}$. This expression is valid only
for $a\le\case{1}{2}$ since $x^{\rm s}_{\rm m}$ must be non-negative.
When $a>a_{\rm c}=\case{1}{2}$, $x^{\rm s}_{\rm m}=0$, i.e., the
critical surface magnetization remains finite for the
semi-infinite system. It follows that our assumption about the
scaling behaviour of $\overline{m}^{\rm d}_{\rm s}$ should be corrected 
in
the regime of first-order transition $(a>\case{1}{2})$. 
It has to compensate the contribution of the product in~\ref{e3.8}, thus
$\overline{m}^{\rm d}_{\rm s}\sim L^{-a}$. Since we are considering the
magnetization of the dual chain, with couplings $J^{\rm
d}_l(a)=1/J_l(a)\simeq J_l(-a)$, this means that when the couplings are
sufficiently weak ($a<-\case{1}{2}$) near $l=1$, the finite-size-scaling
behaviour of the critical magnetization at $l=L$ is anomalous. 

When the perturbation is relevant, $\kappa<1$, the product of the
couplings leads to a stretched exponential behaviour,  
$$
m_{\rm s}\sim\exp\left({a\over1-\kappa}\,L^{1-\kappa}\right)\,,
\label{e3.10}
$$
which is clearly valid only when $a<0$. Otherwise the dual magnetization
term contributes, giving a size-independent result which corresponds to
an ordered surface when $a>0$ \cite{peschel84}. 

\section{Marginal extended surface disorder}

In this section we consider marginal random extended perturbations
at the surface of the TIM. The transverse field is uniform,
$h_l=h=1$, and the distributions of the couplings
$J_l$ are given by either~\ref{e1.2} with $\kappa=\case{1}{2}$
or~\ref{e1.3} with $\kappa=1$ for marginal behaviour. 

{\par\begingroup\parindent=0pt\medskip
\epsfxsize=7truecm
\midinsert
\centerline{\hglue22truemm\epsfbox{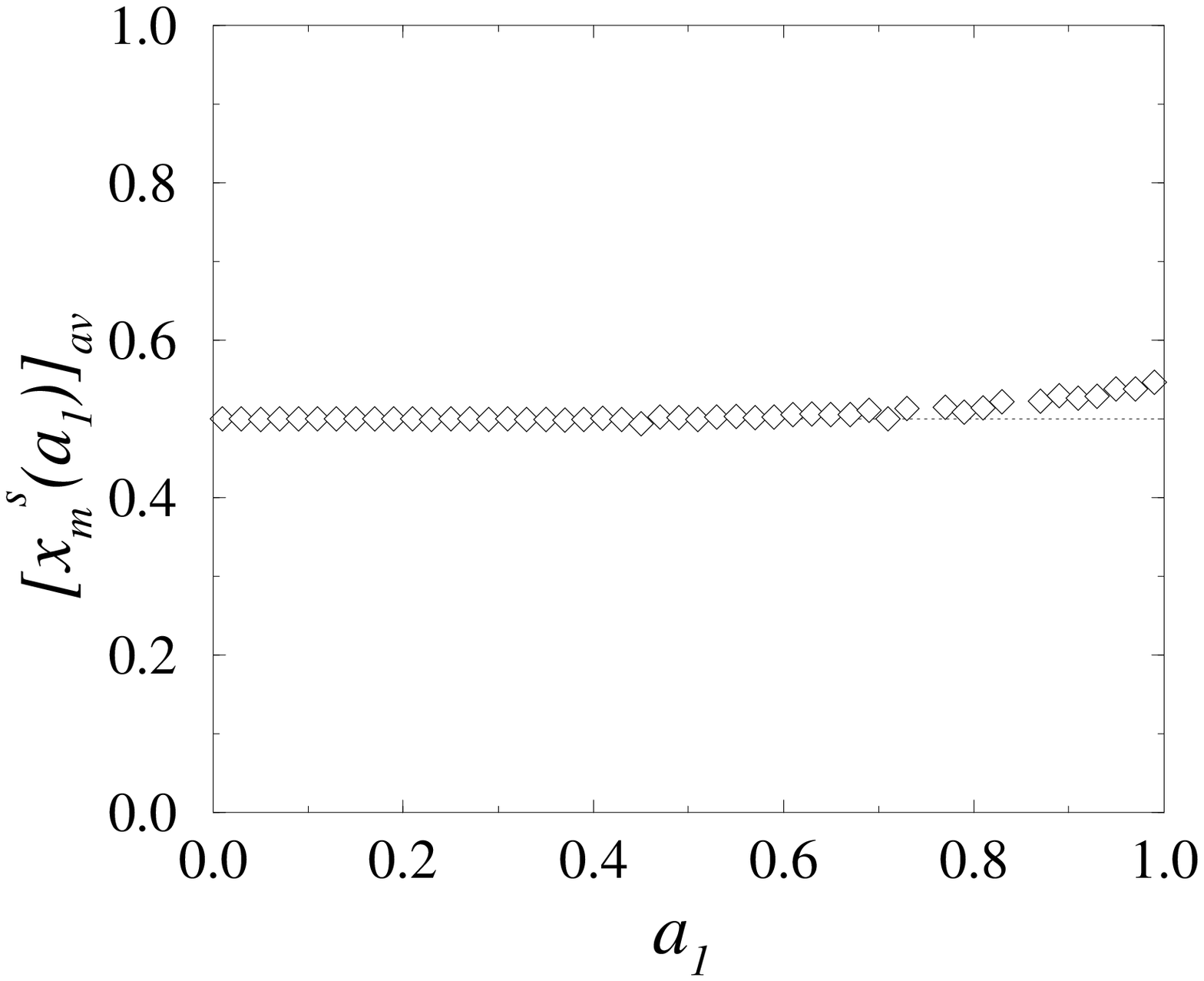}}
\smallskip
\figure{Scaling dimension of the average surface magnetization as a
function of the perturbation amplitude for model \#1. The dotted line
corresponds to the analytical result in
equation~\ref{e4.1}.\label{fig-mod1-beta-av}}  
\medskip
\epsfxsize=7truecm
\centerline{\hglue22truemm\epsfbox{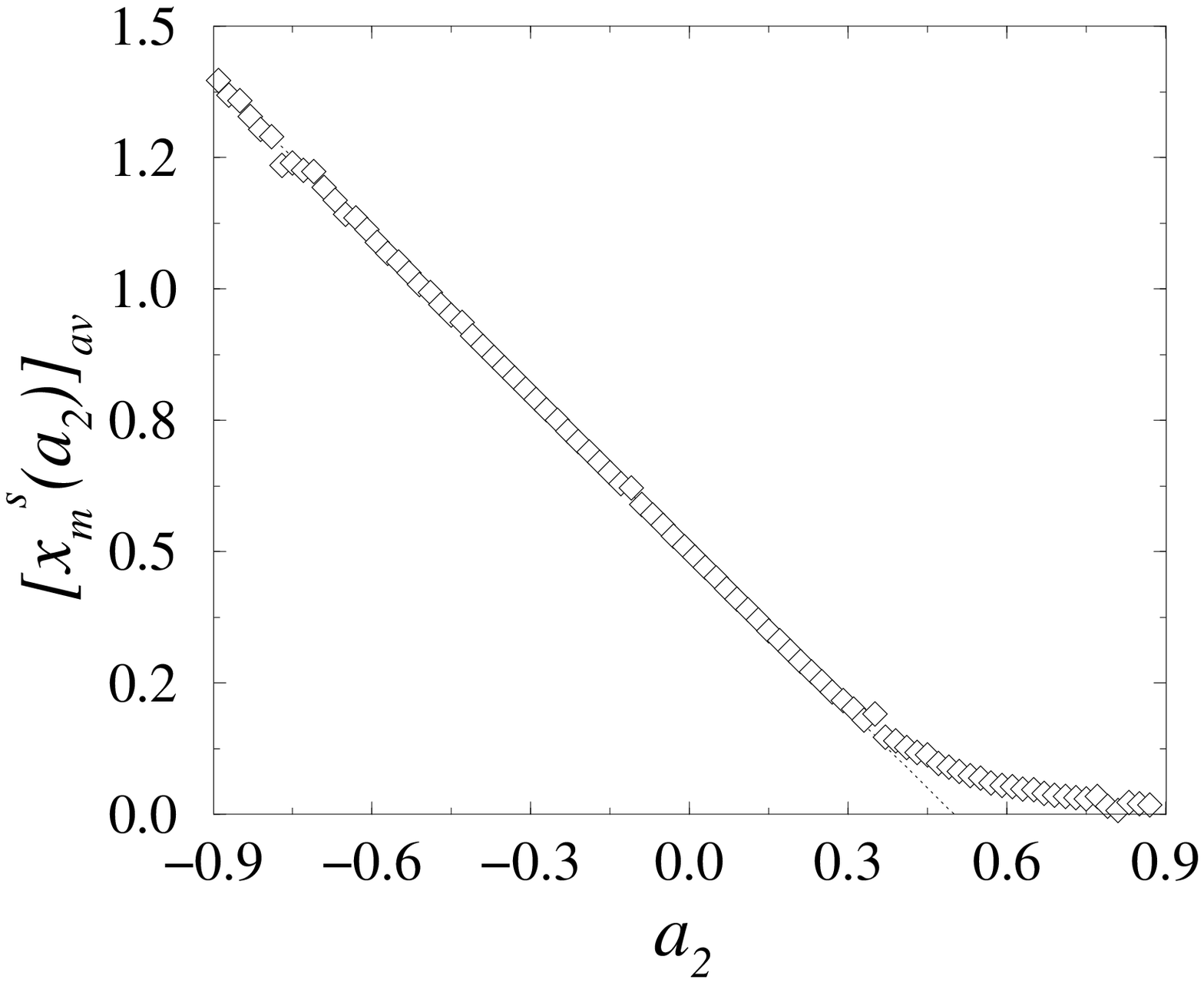}}
\smallskip
\figure{Scaling dimension of the average surface magnetization as a
function of the perturbation amplitude for model \#2. The dotted line
corresponds to the analytical result in
equation~\ref{e4.1}.\label{fig-mod2-beta-av}}  
\endinsert 
\endgroup \par}

The finite-size-scaling behaviour of the {\it average} critical surface
magnetization at $J=1$ is deduced for both models from
equation~\ref{e3.8} for $L\gg1$. As for the HvL model, one expects
$\overline{m}^{\rm d}_{\rm s}$ on the unperturbed end to scale as
$L^{-{1\over2}}$, as long as the transition is continuous, while the
average of the product behaves as $[\prod_{l=1}^L J_l]_{\rm
av}=\prod_{l=1}^L[J_l]_{\rm av}$. 

For model \#1, $[J_l]_{\rm av}=J=1$ and $[m_{\rm s}]_{\rm av}\sim
L^{-{1\over2}}$, whereas $\prod_{l=1}^L[J_l]_{\rm av}\sim L^{a_2}$ for
model \#2. Thus we obtain the following values for the scaling dimension
of the average surface magnetization: 
$$ 
\fl
[x^{\rm s}_{\rm m}]_{\rm av}={1\over2}
\quad{\rm (model\ \#1)}\,,\qquad 
\left\{\matrix{
[x^{\rm s}_{\rm m}]_{\rm av}=1/2-a_2&\quad a_2\le1/2\cr
[x^{\rm s}_{\rm m}]_{\rm av}=0\hfill&\quad a_2>1/2\cr
}
\right.\quad{\rm (model\ \#2)}\,. 
\label{e4.1}
$$
As for the HvL model there is a finite average surface
magnetization at the critical point for strong enough local
couplings, i.e., for $a_2>\case{1}{2}$ where
$\lim_{L\to\infty}[m_{\rm s}]_{\rm av}>0$. In this first-order
regime the dual magnetization no longer scales as $L^{-1/2}$. Actually
the correlation between the two factors on the r.h.s. of
equation~\ref{e3.8} modifies the scaling behaviour of the average of the
product, leading to $[x^{\rm s}_{\rm m}]_{\rm av}=0$. One may notice 
that
since the bulk Ising system has a correlation length exponent
$\nu=1$, equation~\ref{e4.1} also gives the surface magnetic exponent
$[\beta_{\rm s}]_{\rm av}=[x^{\rm s}_{\rm m}]_{\rm av}$.   

The results for the average surface magnetization have been
checked numerically, considering $10^6$ realizations of randomly
inhomogeneous quantum spin chains and computing the critical
magnetization $m_{\rm s}(L)$ given by~\ref{e3.7} for chain sizes
$L=2^5$ to $2^{14}$. The exponents were deduced from an extrapolation
of two-point approximants using the BST algorithm \cite{henkel88}. The
variations of the average surface magnetic exponents, shown in
figures~\ref{fig-mod1-beta-av} and~\ref{fig-mod2-beta-av}, are in
excellent agreement with the expected ones. 

{\par\begingroup\parindent=0pt\medskip
\epsfxsize=7truecm
\topinsert
\centerline{\hglue22truemm \epsfbox{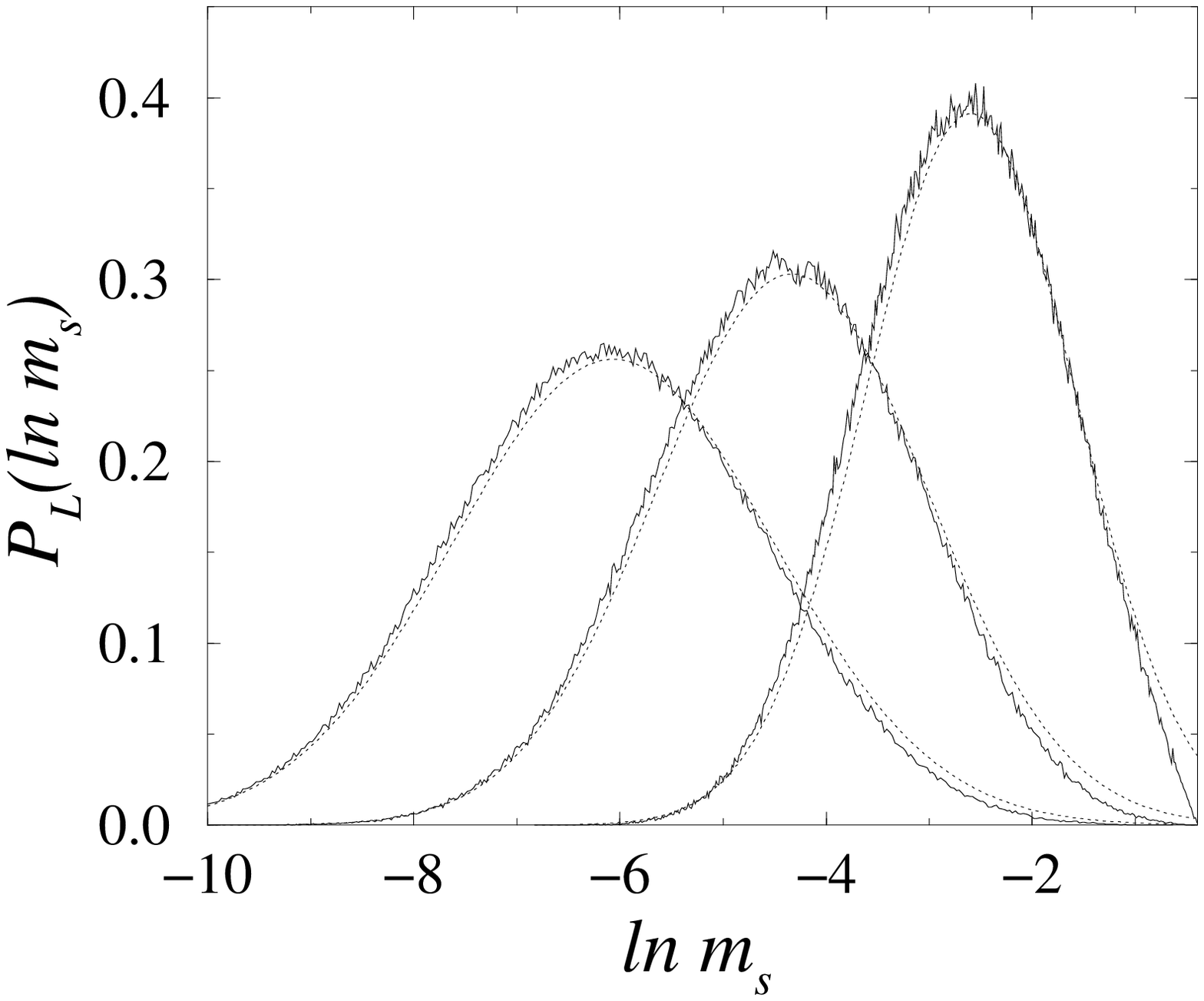}}
\smallskip
\figure{Probability distribution of $\ln m_{\rm s}$
calculated on $10^6$ samples for model \#1. The perturbation amplitude
is $a_1=0.5$ and the chain sizes are $L=2^6$, $2^8$, $2^{10}$, $2^{12}$
and $2^{14}$ from right to left. The dotted line corresponds to the
normal distribution in equation~\ref{e4.2}.\label{fig-mod1-plnms}}  
\endinsert 
\endgroup \par}

Next we are going to study the probability distribution of the
surface magnetization for the two models. Again we start from
equation~\ref{e3.8} in which, for large $L$, 
$\overline{m}^{\rm d}_{\rm s}$, which refers to the unperturbed end,
is expected to scale as $L^{-{1\over2}}$ as long as the transition on 
the
other surface is second order. Thus, in the second-order regime, the
central limit theorem can be applied to the random variable $\ln m_{\rm
s}+\case{1}{2}\ln L=\sum_{l=1}^L\ln(J_l)$ and $\ln m_{\rm s}$ follows a
normal distribution:  
$$ 
P_L(\ln m_{\rm s})\simeq{1\over\sqrt{2\pi\,{\rm var}(\ln
m_{\rm s})}} \exp\left[-{(\ln m_{\rm s}-[\ln m_{\rm s}]_{\rm av})^2\over
2\,{\rm var}(\ln m_{\rm s})}\right]\,, 
\label{e4.2}
$$
where ${\rm var}(\ln m_{\rm s})=\left[(\ln m_{\rm s}-[\ln m_{\rm 
s}]_{\rm
av})^2\right]_{\rm av}$. This expression is valid for $L\gg1$ and $\ln
m_{\rm s}\ll0$. It does not correctly describe the rare events,
governing the average behaviour, which correspond to the tail of
the distribution near $\ln m_{\rm s}=0$. 

The parameters of the distribution are easily deduced from
equations~\ref{e1.2} and~\ref{e1.3} and read:
$$
\fl
\eqalign{
&\left\{\eqalign{
                 &[\ln m_{\rm s}]_{\rm av}\simeq-{1\over2}\ln
L-\sum_{l=1}^L{a_1^2\over2l}\simeq-{1\over2}(1+a_1^2)\ln L\,,\cr 
                 &{\rm var}(\ln m_{\rm s})\simeq a_1^2\ln L\,.\cr
                 }
\right.\qquad{\rm (model\ \#1)}\cr
&\left\{\eqalign{
                &[\ln m_{\rm s}]_{\rm av}
\simeq-{1\over2}\ln L+\sum_{l=1}^L{\ln(1+a_2)\over l}
\simeq-\left[{1\over2}-\ln(1+a_2)\right]\ln L\,,\cr  
                &{\rm var}(\ln m_{\rm s})\simeq \ln^2(1+a_2)\ln L\,.\cr
                }
\right.\qquad{\rm (model\ \#2)}\cr
}
\label{e4.3}
$$

{\par\begingroup\parindent=0pt\medskip
\epsfxsize=7truecm
\topinsert
\centerline{\hglue22truemm \epsfbox{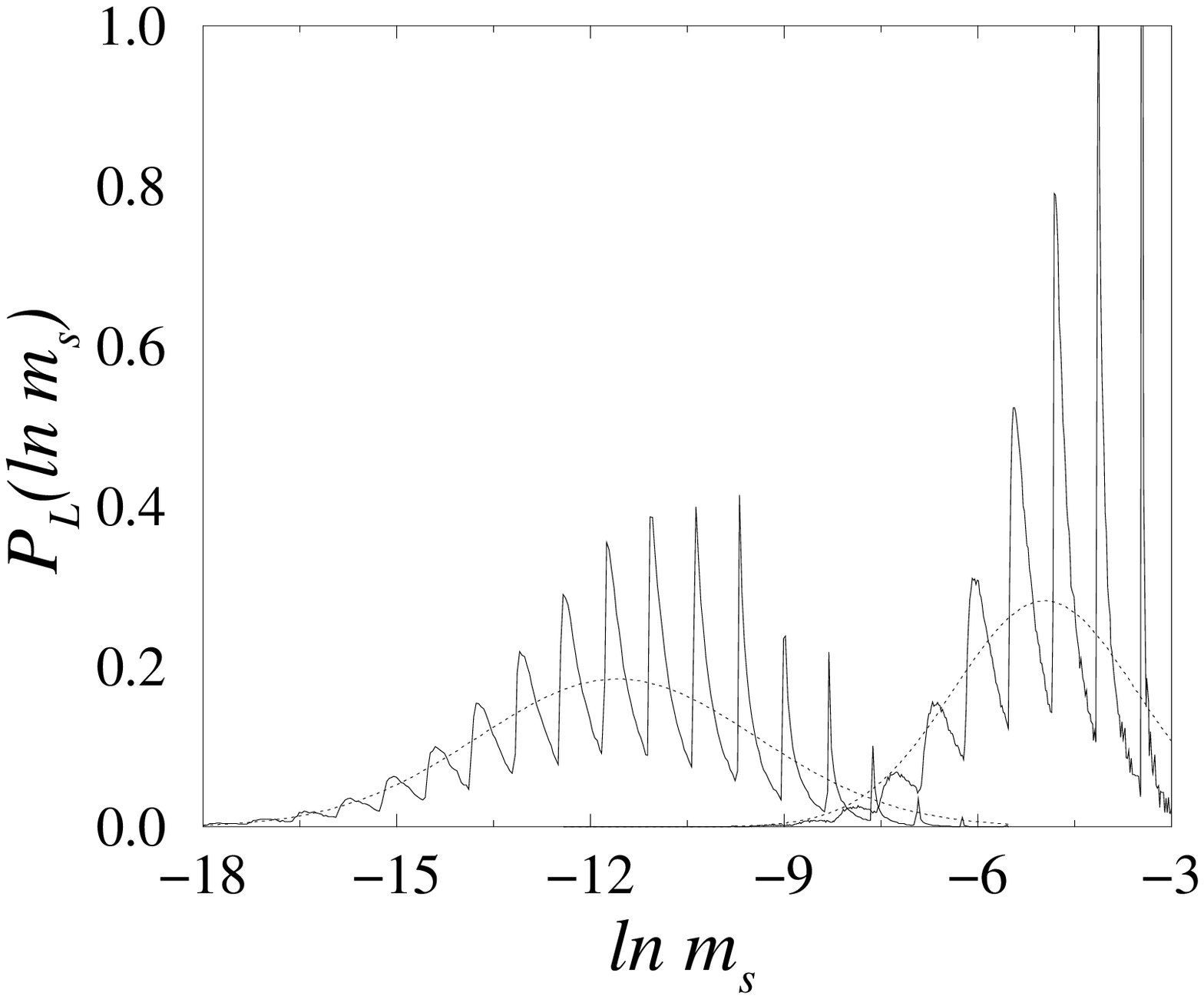}}
\smallskip
\figure{Probability distribution of $\ln m_{\rm s}$
calculated on $10^6$ samples for model \#2 in the second-order regime.
The perturbation amplitude is $a_2=-0.5$ and the chain sizes are $L=2^6$
and $2^{14}$ from right to left. The dotted line corresponds to the 
normal
distribution in equation~\ref{e4.2}. The actual distribution displays
strong periodic oscillations around the normal distribution. 
\label{fig-mod2-plnms}}     
\endinsert 
\endgroup \par}

{\par\begingroup\parindent=0pt\medskip
\epsfxsize=7truecm
\midinsert
\centerline{\hglue22truemm \epsfbox{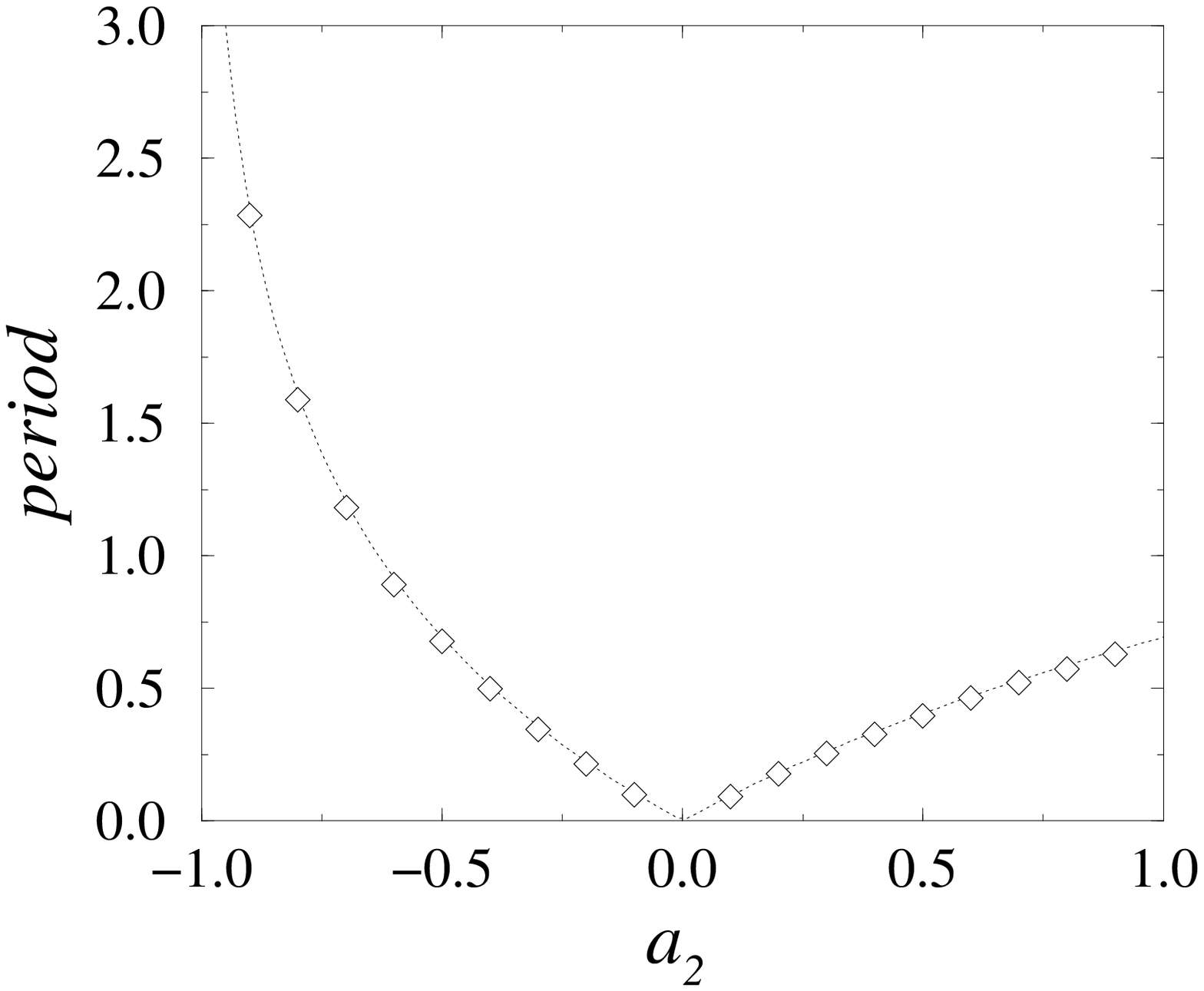}}
\smallskip
\figure{Variation of the period of oscillations in the distribution
$P_L(\ln m_{\rm s})$ for model \#2 as a function of the
marginal amplitude $a_2$. The points give the periods
deduced from the numerical data for the largest size, $L=2^{14}$, and 
the 
dotted line corresponds to $\vert\ln(1+a_2)\vert$.
\label{fig-mod2-period}}    
\endinsert 
\endgroup \par}

The distributions $P_L(\ln m_{\rm s})$, obtained through
numerical calculations on $10^6$ samples for different chain sizes, are
compared to the analytical expressions in figures~\ref{fig-mod1-plnms}
and~\ref{fig-mod2-plnms}. The agreement is quite good at large sizes,
although the normal distribution is strongly modulated for model \#2. 
As shown in figure~\ref{fig-mod2-period}, the modulation involves a
periodic function of period $1$ in the variable $\ln m_{\rm
s}/\vert\ln(1+a_2)\vert$, where $1+a_2$ is the ratio of the two
couplings in equation~\ref{e1.3}. The oscillations are log-periodic in
the variable $m_{\rm s}$ (see reference~\cite{sornette98} for a recent
review about log-periodic phenomena).  

{\par\begingroup\parindent=0pt\medskip
\epsfxsize=7truecm
\topinsert
\centerline{\hglue22truemm \epsfbox{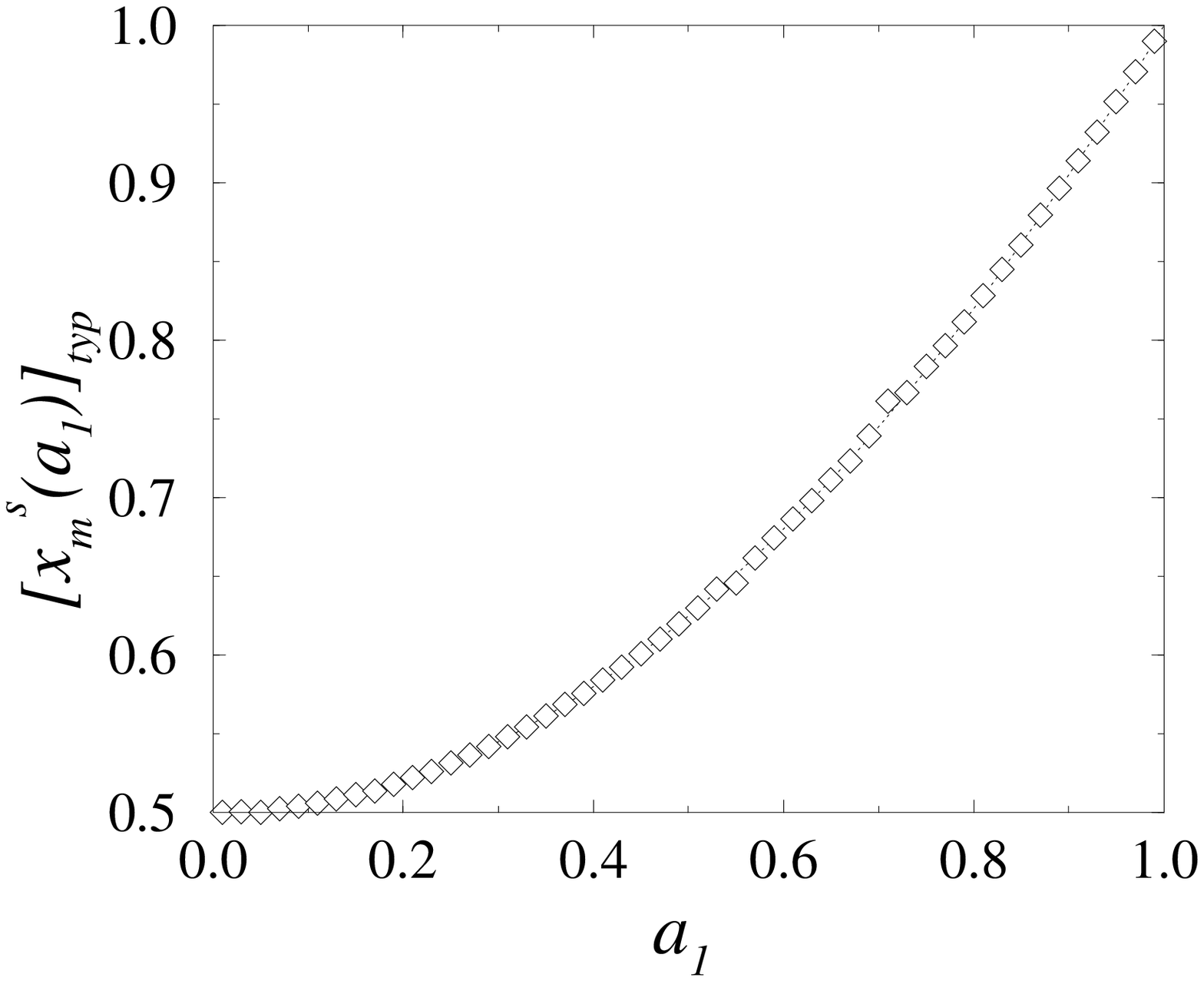}}
\smallskip
\figure{Scaling dimension of the typical surface magnetization as a
function of the perturbation amplitude $a_1$ for model \#1. The dotted
line corresponds to the analytical result in
equation~\ref{e4.4}.\label{fig-mod1-beta-typ}}  
\endinsert 
\endgroup \par}

{\par\begingroup\parindent=0pt\medskip
\epsfxsize=7truecm
\midinsert
\centerline{\hglue22truemm \epsfbox{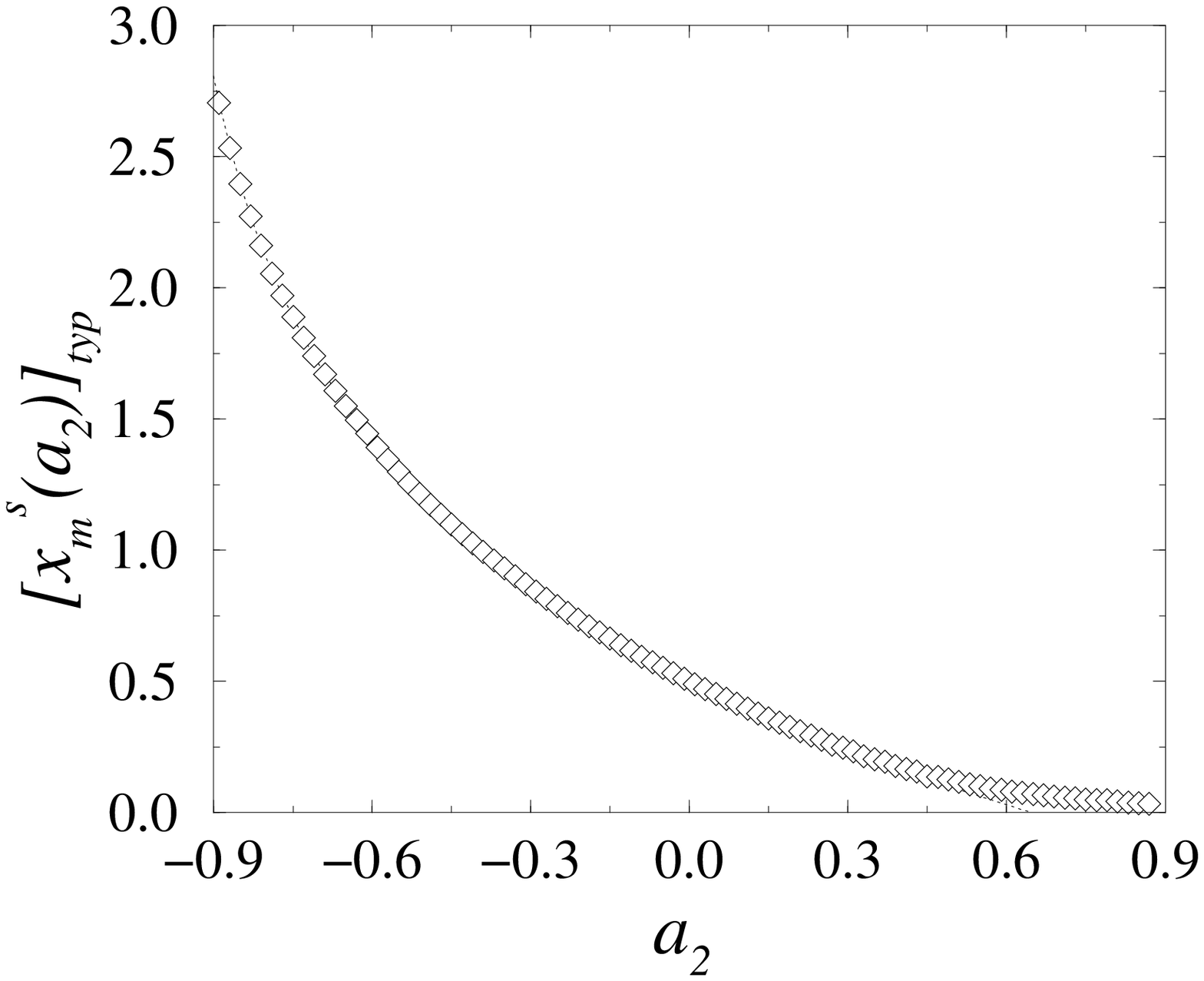}}
\smallskip
\figure{Scaling dimension of the typical surface magnetization as a
function of the perturbation amplitude for model \#2. The dotted line
corresponds to the analytical result in
equation~\ref{e4.4}.\label{fig-mod2-beta-typ}}  
\endinsert 
\endgroup \par}
 
The probability distribution is centered on $[\ln m_{\rm s}]_{\rm
av}=\ln [m_{\rm s}]_{\rm typ}\sim \ln L$, where $[m_{\rm s}]_{\rm typ}$
is the most probable or {\it typical} value of the surface
magnetization. Consequently, the typical magnetization scales as a power
of $L$, $[m_{\rm s}]_{\rm typ}\sim  L^{-[x^{\rm s}_{\rm m}]_{\rm typ}}$,
with a continuously varying typical exponent given by: 
$$ 
\eqalign{
&\,\quad[x^{\rm s}_{\rm m}]_{\rm typ}={1\over2}(1+a_1^2)
\qquad{\rm (model\ \#1)}\,,\cr
&\left\{\matrix{
[x^{\rm s}_{\rm m}]_{\rm typ}=1/2-\ln(1+a_2)
&\quad a_2\le\sqrt{\e}-1\cr
[x^{\rm s}_{\rm m}]_{\rm typ}=0\hfill
&\quad a_2>\sqrt{\e}-1\cr
}
\right.\qquad{\rm (model\ \#2)}\,.\cr
}
\label{e4.4}
$$

In model \#1, the typical surface magnetization vanishes continuously at
the bulk critical point for any value of the perturbation amplitude. In
model \#2, for strong enough local couplings, $a_2>[a_{2{\rm c}}]_{\rm
typ}=\sqrt{\e}-1\simeq0.6487$, there is a first-order surface
transition,  the typical surface magnetization remaining nonvanishing at
criticality. In the range $\case{1}{2}< a_2<\sqrt{\e}-1$, at the bulk
critical point, $[m_{\rm s}]_{\rm av}>0$ whereas $[m_{\rm s}]_{\rm
typ}=0$ when $L\to\infty$, a behaviour which is consistent with the fact
that $[\ln m_{\rm s}]_{\rm av}\le\ln [m_{\rm s}]_{\rm av}$. 

The numerical results for the scaling dimension of the typical surface
magnetization are shown in figures~\ref{fig-mod1-beta-typ}
and~\ref{fig-mod2-beta-typ}. Two-point finite-size approximants were
extrapolated using the BST algorithm and we used the same samples as for
the average magnetization. The agreement with the expressions given
in~\ref{e4.4} is quite good.

\section{Relevant extended surface disorder}

Let us now examine the modified surface critical behaviour induced by
relevant extended surface disorder which, according to the disussion of
section 2, corresponds to a decay exponent $\kappa$ in
equations~\ref{e1.2} and~\ref{e1.3} lower than $\kappa_1=1/2$ for model
\#1 and $\kappa_2=1$ for model \#2, respectively. 

Like in the marginal case, the average surface
magnetization obtained by averaging equation~\ref{e3.8} would scale
with an exponent $[x_{\rm m}^{\rm s}]_{\rm av}=1/2$ for model \#1 if the
dual magnetization was unaffected by the
perturbation. A numerical finite-size scaling study with
$a_1=0.5$ and $\kappa=0.25$, shows that the actual behaviour is
a stretched exponential one since $\ln [m_{\rm s}]_{\rm av}\sim 
L^{1/2}$,
as shown in figure~\ref{fig-mod12-rel-lnms}. Thus the slowly decaying
perturbation  changes the critical behaviour on the second surface. By
symmetry, the behaviour is the same for negative values of $a_1$.
Numerical studies for other values of $\kappa$ indicate that the
argument of the stretched exponential involves the power
$L^{1-2\kappa}$. 

{\par\begingroup\parindent=0pt\medskip
\epsfxsize=7truecm
\midinsert
\centerline{\hglue22truemm \epsfbox{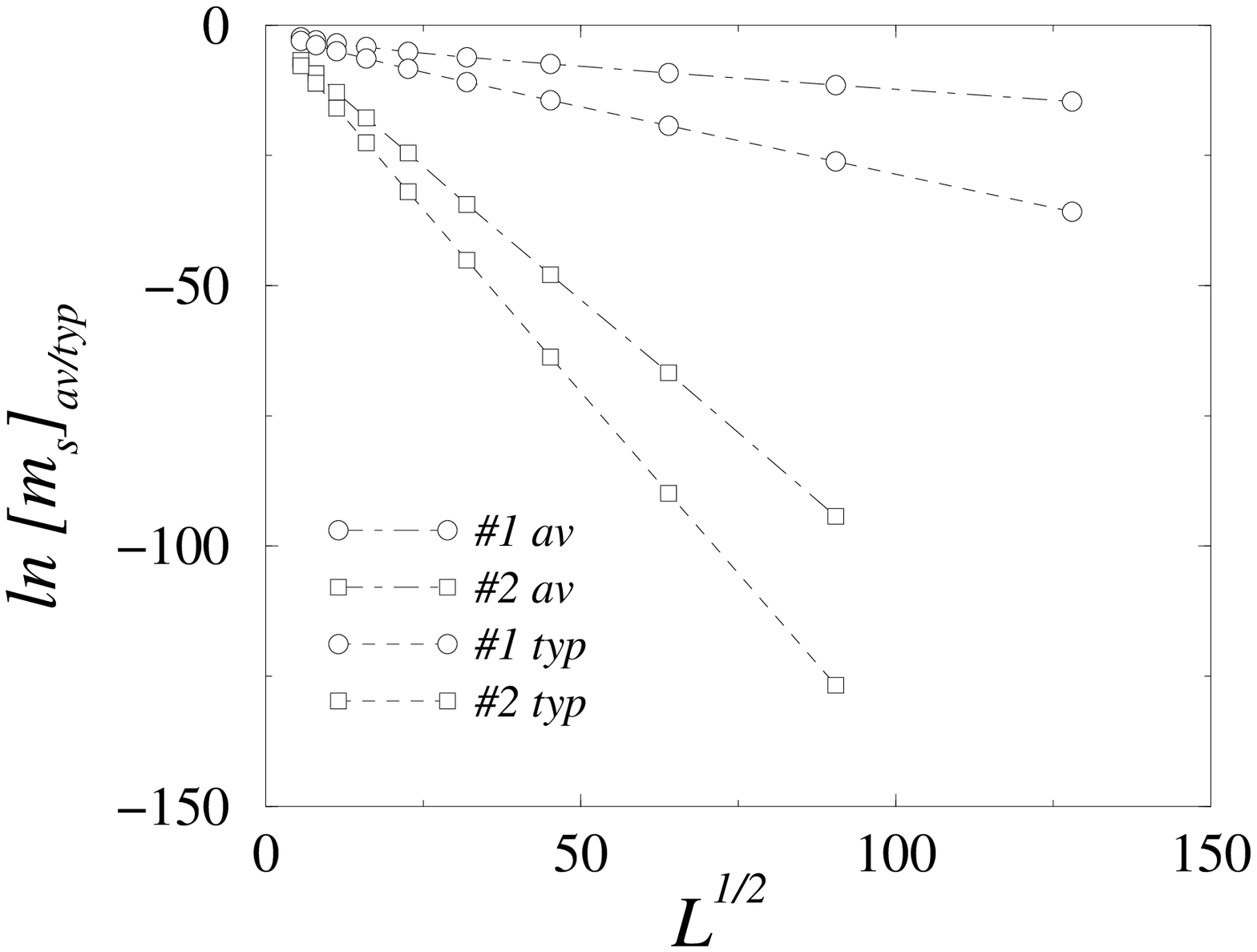}}
\smallskip
\figure{Finite-size scaling of the average (dot-dashed lines) and
typical (dashed lines) surface magnetization with relevant perturbations
illustrating the stretched exponential behaviour. The parameters are 
$a_1=0.5$ and $\kappa=0.25$ for model \#1 (circles), $a_2=-0.5$ and
$\kappa=0.5$ for model \#2 (squares). Averages were taken
over $10^6$ samples.
\label{fig-mod12-rel-lnms}}      
\endinsert   
\endgroup 
\par}

{\par\begingroup\parindent=0pt\medskip
\epsfxsize=7truecm
\midinsert
\centerline{\hglue22truemm \epsfbox{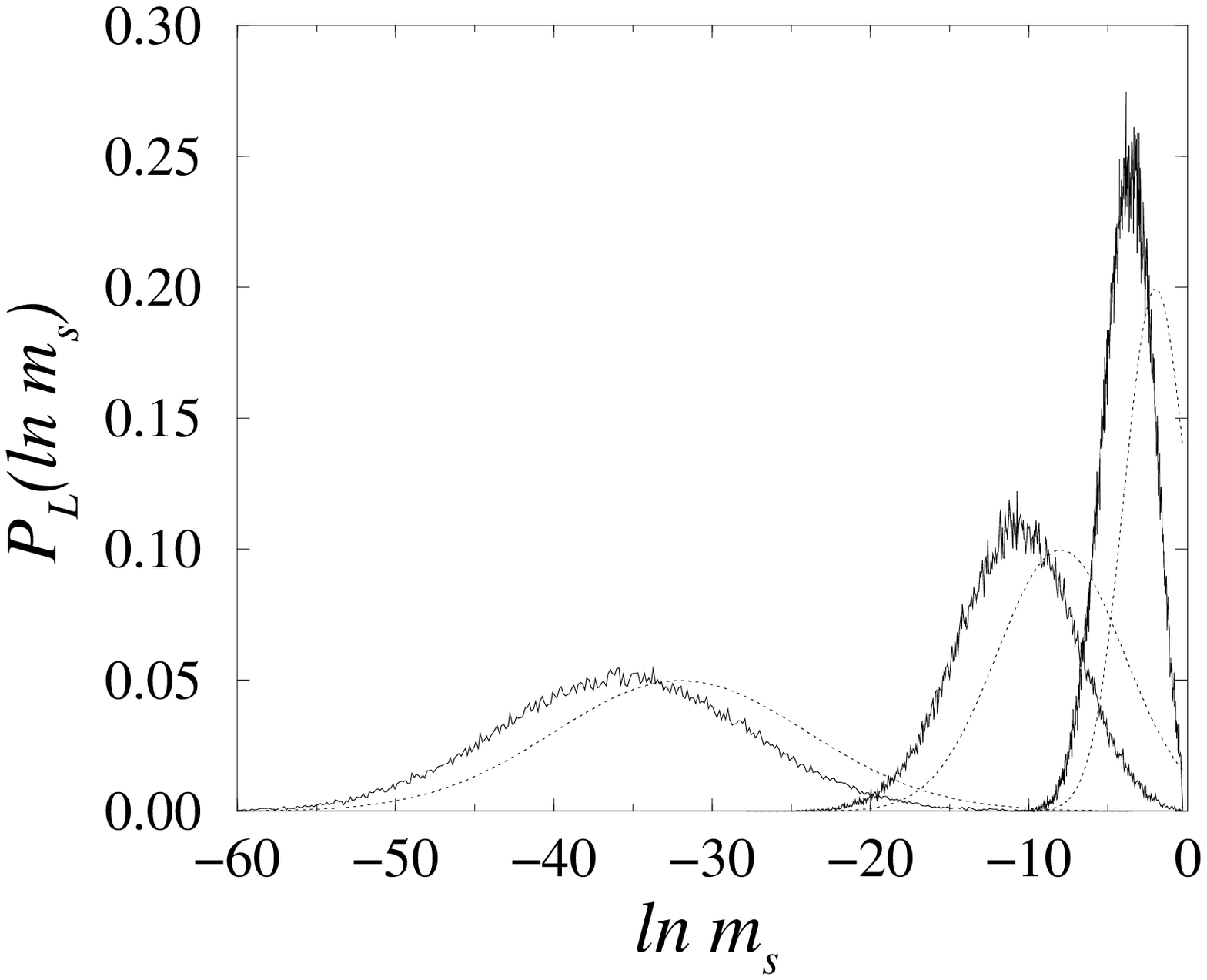}}
\smallskip
\figure{Probability distribution of $\ln m_{\rm s}$ calculated on $10^5$
samples for model \#1 with a relevant perturbation. The 
amplitude is $a_1=0.5$ and the decay exponent $\kappa=0.25$. The chain
sizes are $L=2^6$, $2^{10}$ and $2^{14}$ from right to left. The dotted
lines correspond to the normal distribution in equation~\ref{e4.2} when
the contribution of the dual magnetization to $\ln m_{\rm s}$ is
neglected.  \label{fig-mod1-rel-plnms}}   
\endinsert  
\endgroup 
\par} 

For model \#2, taking only into account the
product of the couplings, one obtains:
$$
[m_{\rm s}]_{\rm av}\sim\exp\left[{a_2\over1-\kappa}\,
L^{1-\kappa}\right]\,.
\label{e5.1}
$$
Hence the average surface magnetization has a stretched
exponential dependence on $L$ for negative values of the amplitude 
$a_2$.
The numerical data are in agreement with this behaviour as shown in
figure~\ref{fig-mod12-rel-lnms} for $a_2=-0.5$ and $\kappa=0.5$.
The value of the slope, $-0.99\pm0.04$, is close to
the analytical result, $-1$. Thus the dual magnetization is likely
to contribute here only through a power of $L$ in front of the
exponential.      
 
When $a_2>0$, the divergence of $[m_{\rm s}]_{\rm av}$ with the
system size signals that the dual magnetization 
term in~\ref{e3.8} can no longer be neglected. The surface transition is
then first order. This was checked numerically at $a_2=0.5$ where the
average critical surface magnetization is indeed size-independent at 
large
$L$ values. Contrary to the case of a marginal perturbation where a
mininum enhancement of the couplings was needed to maintain surface
order at the bulk critical point, here a first-order surface transition
occurs for any positive value of the perturbation amplitude. The same
behaviour is obtained with the HvL model~\cite{peschel84}.

{\par\begingroup\parindent=0pt\medskip
\epsfxsize=7truecm
\topinsert
\centerline{\hglue22truemm \epsfbox{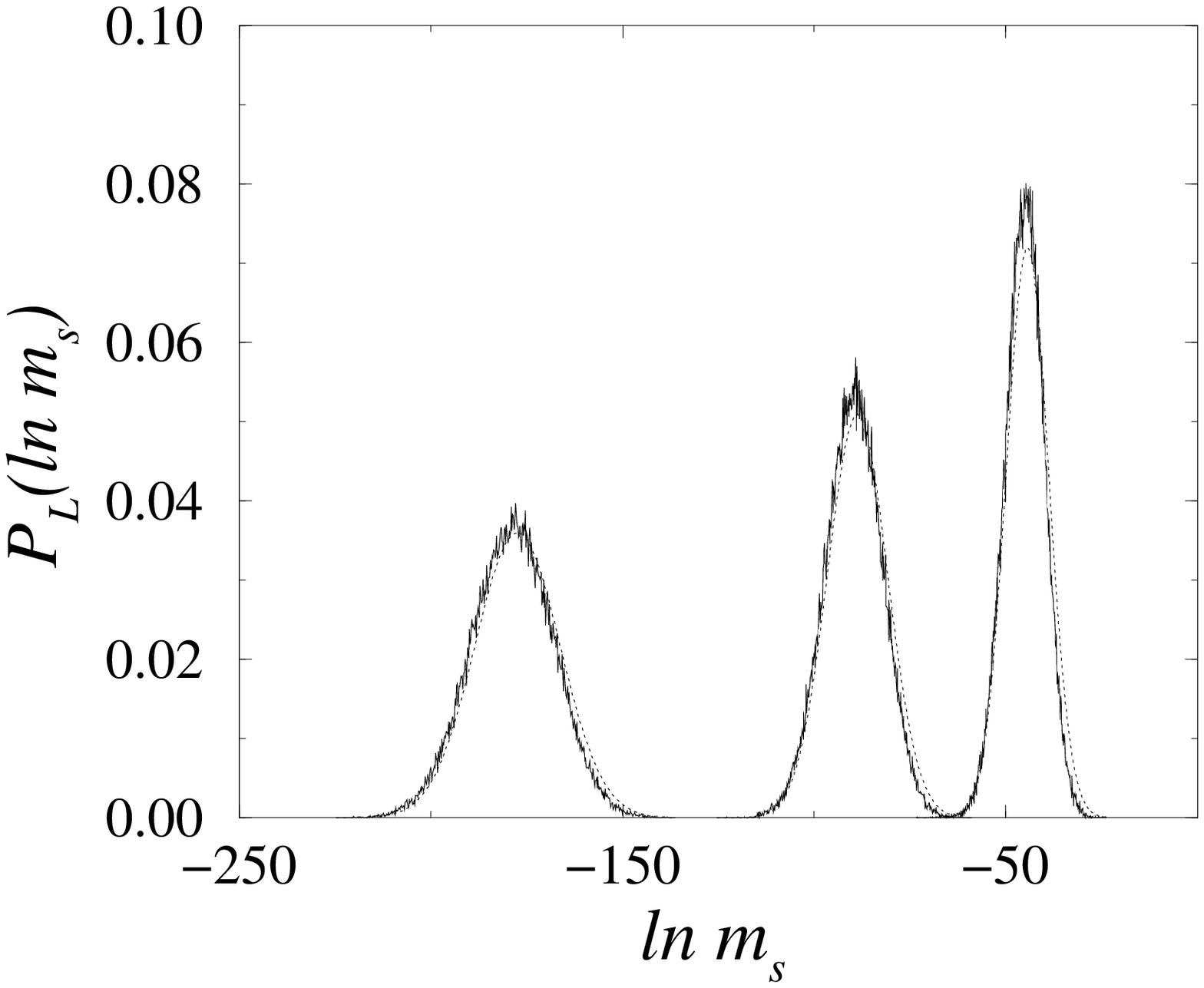}}
\smallskip
\figure{Probability distribution of $\ln m_{\rm s}$ calculated on $10^5$
samples for model \#2 with a relevant perturbation. The 
amplitude is $a_2=-0.5$ and the decay exponent $\kappa=0.5$. The chain
sizes are $L=2^{10}$, $2^{12}$ and $2^{14}$ from right to left. The
dotted lines correspond to the normal distribution in 
equation~\ref{e4.2}
when the contribution of the dual magnetization to $\ln m_{\rm s}$ is
neglected.\label{fig-mod2-rel-plnms}}   
\endinsert 
\endgroup \par}

The probability distribution $P_L(\ln m_{\rm s})$ can be obtained in the
same way as in the marginal situation when the transition is continuous,
provided that $\ln m_{\rm s}$ is dominated by the contribution of the
product in~\ref{e3.8}. Then one recovers the normal distribution of
equation~\ref{e4.2} with the following parameters:
$$
\fl
\eqalign{
&\left\{\eqalign{
                 &[\ln m_{\rm s}]_{\rm av}
\simeq-\sum_{l=1}^L{a_1^2\over 2l^{2\kappa}}
\simeq-{a_1^2\over2(1-2\kappa)}\, L^{1-2\kappa}\,,\cr 
                 &{\rm var}(\ln m_{\rm s})
\simeq {a_1^2\over1-2\kappa}\, L^{1-2\kappa}\,.\cr
                 }
\right.\qquad\quad{\rm (model\ \#1)}\cr
&\left\{\eqalign{
                &[\ln m_{\rm s}]_{\rm av}
\simeq\sum_{l=1}^L{\ln(1+a_2)\over l^\kappa}
\simeq{\ln(1+a_2)\over1-\kappa}\, L^{1-\kappa}\,,\cr  
                &{\rm var}(\ln m_{\rm s})
\simeq{\ln^2(1+a_2)\over1-\kappa}\, L^{1-\kappa}\,.\cr
                }
\right.\qquad{\rm (model\ \#2)}\cr
}
\label{e5.2}
$$
The numerical data, shown in figures~\ref{fig-mod1-rel-plnms}
and~\ref{fig-mod2-rel-plnms}, were deduced from $10^5$ realizations and,
as above, with $a_1=0.5$, $\kappa=0.25$ for model \#1 and $a_2=-0.5$,
$\kappa=0.5$ for model \#2. 

In the case of model \#1, the shift with respect to the analytical 
results
confirms a weak relevant contribution of the dual magnetization term
in~\ref{e3.8}. This is confirmed in figure~\ref{fig-mod12-rel-lnms} 
where
the slope of $\ln [m_{\rm s}]_{\rm typ}$ versus $L^{1-2\kappa}$ is
$0.266\pm0.003$ to be compared to the value $-0.25$ given by~\ref{e5.2}. 

The agreement is much better for model \#2. Here the slope of $\ln
[m_{\rm s}]_{\rm typ}$ versus $L^{1-\kappa}$ in
figure~\ref{fig-mod12-rel-lnms}, $-1.401\pm0.002$, has a smaller
relative deviation from the expected value, $-1.386$. One may notice 
that
the probability distribution no longer displays the strong log-periodic
modulations observed in the marginal case. 

\section{Discussion}

The two models of extended surface disorder studied in this paper lead
to a surface critical behaviour which is quite similar to that
obtained in the HvL model, with continuously varying exponents in the
marginal case and a stretched exponential behaviour for relevant
perturbations. The main differences are the non-self-averaging
behaviour of the random models and the the log-periodicy obtained with 
model \#2.

This log-periodic behaviour is easily understood if one considers the 
expression of the surface magnetization in equation~\ref{e3.8}. At 
criticality $h_l=J=1$ and, according to~\ref{e1.3}, $\ln J_l$ is either 
0 for unperturbed couplings or $\ln(1+a_2)$ for perturbed ones. A sample 
$\alpha$ with $n_\alpha$ modified couplings contributes to the 
probability distribution $P_L(\ln m_{\rm s})$ at a value 
$n_\alpha\ln(1+a_2)+\ln\overline{m}^{\rm d}_{{\rm s}\alpha}$ of $\ln 
m_{\rm s}$. If the fluctuations of the last term from sample to sample  
are sufficiently weak compared to the interval $\ln(1+a_2)$ a modulation 
with period $\ln(1+a_2)$ is obtained for $P_L(\ln m_{\rm s})$. The 
argument does not hold for model \#1 since the values of the couplings 
vary from site to site. The absence of modulation for model \#2 with a 
relevant perturbation gives a lower bound for the influence of the 
disorder on the second surface.

The connection to the HvL model can be clarified through the
introduction of effective HvL models describing either the average or
the typical behaviour. As long as the size dependence of the dual
magnetization in equation~\ref{e3.8} is the same as for an unperturbed
surface, one may introduce effective HvL interactions such that:
$$
[J_l]^{\rm eff}_{\rm av}=[J_l]_{\rm av}\,,\qquad
[J_l]^{\rm eff}_{\rm typ}=\exp([\ln J_l]_{\rm av})\,.
\label{e6.1}
$$  
Equations~\ref{e1.2} and~\ref{e1.3} then lead to:
$$
\eqalign{
&\left\{
\eqalign{
&[J_l]^{\rm eff}_{\rm av}=J\,,\cr
&[J_l]^{\rm eff}_{\rm typ}\simeq 
J\left(1-{a_1^2\over 2l^{2\kappa}}\right)\,.\cr
}
\right.\qquad{\rm (model\ \#1)}\cr
&\left\{
\eqalign{
&[J_l]^{\rm eff}_{\rm av}=J\left(1+{a_2\over l^\kappa}\right)\,,\cr
&[J_l]^{\rm eff}_{\rm typ}\simeq
J\left[1+{\ln(1+a_2)\over l^{\kappa}}\right]\,.\cr
}
\right.\qquad{\rm (model \#2)}\cr
}
\label{e6.2} 
$$
For model \#1, $\kappa_{\rm HvL}=2\kappa$ and the
random perturbation is marginal at $\kappa_1=1/2$. The effective
amplitude vanishes for averaged quantities whereas it is always
negative, with $[a]^{\rm eff}_{\rm typ}=-a_1^2/2$, for typical 
quantities. Hence the surface transition is always continuous and the
marginal exponents in~\ref{e4.1} and~\ref{e4.4} follow from~\ref{e3.9}
with the appropriate effective amplitude. 

For model \#2, $\kappa_{\rm HvL}=\kappa$ leads to $\kappa_2=1$. The
effective amplitudes are  $[a]^{\rm eff}_{\rm av}=a_2$ and $[a]^{\rm
eff}_{\rm typ}=\ln(1+a_2)$. The continuously varying exponent can be
deduced from the HvL results when the transition is continuous, in
this case even for relevant perturbations (compare~\ref{e5.1}, 
\ref{e5.2}
and~\ref{e3.10}).

One may go further and conjecture the form of the scaling dimension of 
the
surface energy density in the second-order regime. Its value $x^{\rm
s}_{\rm e}=2(1-a)$ for the HvL model \cite{igloi93} translates into:
$$ 
\eqalign{
&\left\{\matrix{
[x^{\rm s}_{\rm e}]_{\rm av}=2\hfill\cr
[x^{\rm s}_{\rm e}]_{\rm typ}=2+a^2_1\cr
}
\right.\qquad{\rm (model\ \#1)}\,,\cr
&\left\{\matrix{
[x^{\rm s}_{\rm e}]_{\rm av}=2(1-a_2)\hfill
&\quad a_2\le1/2\hfill\cr
[x^{\rm s}_{\rm e}]_{\rm typ}=2[1-\ln(1+a_2)]
&\quad a_2\le\sqrt{\e}-1\cr
}
\right.\qquad{\rm (model\ \#2)}\,.\cr
}
\label{e6.3}
$$

As already mentioned, the inhomogeneously disordered Ising
quantum chain  corresponds to the extreme anisotropic limit of a $2d$
classical Ising model with correlated disorder along the rows parallel 
to its surface. It is easy to apply the considerations of section 2 to
the case where the perturbations of the different bonds (or sites) at a distance
$l$ from the surface of a $d$-dimensional classical system are independent random
variables distributed according to~\ref{e1.2} or~\ref{e1.3}, i.e., with
$d$-dimensional inhomogeneous disorder. 

For model \#2, to first order, the perturbation amplitude $A_2$ transforms like
in~\ref{e2.9} with a scaling dimension $y_\phi-\kappa$ when it couples to the
field $\phi(\bi{r})$ in $d$ dimensions. The off-diagonal
second-order correction has the same behaviour. The diagonal second-order
correction, with scaling dimension $2y_\phi-\kappa-d$,
cannot be more relevant than the first-order correction since $y_\phi\leq d$.
Thus the relevance of the perturbation remains the same as for the HvL model.

For model \#1, however, the scaling dimension of the second-order
perturbation amplitude $A_1^2$ in equation~\ref{e2.6} changes to
$y_{A_1^2}=2y_\phi-2\kappa-d$. Thus a thermal perturbation is
marginal when $\kappa=1/\nu-d/2=\alpha/(2\nu)$ where $\alpha$ is the bulk
specific heat exponent. 
For the $2d$ Ising model, with $\alpha=0$, the perturbation is
irrelevant as soon as $\kappa>0$ and marginal for homogeneous disorder, 
a well-known result \cite{harris74}. The $2d$ $q$-state Potts model would 
be more interesting since marginal behaviour then corresponds to
$\kappa=1/5$ for $q=3$ ($\nu=5/6$) and $\kappa=1/2$ for $q=4$ ($\nu=2/3$)
\cite{wu82}. One could then interpolate between a marginal inhomogeneous
perturbation and a relevant homogeneous one. Another case of interest is
the $2d$ $q$-state Potts model with $q>4$. Here the pure system
has a first-order bulk transition \cite{baxter73,baxter82}, the surface
transition is continuous \cite{lipowsky82,igloi99b} and homogeneous
disorder changes the bulk transition from first to second order
[30--32]. With $\nu=1/2$ at the
discontinuity fixed point \cite{nienhuis75,fisher82}, the critical
decay exponent for marginal behaviour is $\kappa=1$. Varying $\kappa$ 
from 1 to 0 could lead to interesting new phenomena.

To conclude, let us mention that it would be useful to adapt Fisher's 
renormalization group approach to the type of perturbations considered 
in this work.

\ack This work has been supported by the French-Hungarian
cooperation program Balaton (Minist\`ere des Affaires
Etrang\`eres-O.M.F.B.), the Hungarian National Research
Fund under grants No TO23642 and M 028418 and by the Hungarian Ministery
of Education under grant No FKFP 0765/1997. The Laboratoire de Physique
des Mat\'eriaux is Unit\'e Mixte de Recherche C.N.R.S. No 7556.

\numreferences

\bibitem{rieger97a} {Rieger H and Young A P 1997}\ {\it Complex
Behavior of Glassy Systems (Lecture Notes in Physics vol~492)}\  
ed M Rubi and C Perez-Vicente\ (Heidelberg: Springer-Verlag)\ p~256
\bibitem{fisher92} {Fisher D S 1992}\ {\PRL}\ {\bf 69}\ {534}
\bibitem{fisher95} {Fisher D S 1995}\ {\PR\ \it B}\ {\bf 51}\ {6411}
\bibitem{young96} {Young A P and Rieger H 1996}\ {\PR \it B}\ {\bf 53}\
{8486}
\bibitem{igloi97} {Igl\'oi F and Rieger H 1997}\ {\PRL}\ {\bf 78}\ 
{2473}
\bibitem{young97} {Young A P 1997}\ {\PR \it B}\ {\bf 56}\ {11 691}
\bibitem{rieger97b} {Rieger H and Igl\'oi F 1997}\ {\it Europhys. 
Lett.}\
{\bf 39}\ {135}
\bibitem{igloi98a} {Igl\'oi F and Rieger H 1998}\ {\PR \it B}\ {\bf 57}\
{11 404}
\bibitem{igloi98b} {Igl\'oi F and Rieger H 1998}\ {\PR \it E}\ {\bf 58}\
{4238} 
\bibitem{igloi99a} {Igl\'oi F, Juh\'asz R and Rieger H 1999}\
{\PR \it B}\ {\bf **}\ {**}
\bibitem{hilhorst81} {Hilhorst H J and van Leeuwen J M J
1981}\ {\PRL}\ {\bf 47}\ {1188} 
\bibitem{blote83} {Bl\"ote H W J and Hilhorst H J 1983}\ {\PRL}\ {\bf 
51}\
{20} 
\bibitem{blote85} {Bl\"ote H W J and Hilhorst H J 1985}\ {\JPA}\
{\bf 18}\ {3039}
\bibitem{cordery82} {Cordery R 1982}\ {\PRL}\ {\bf 48}\ {215} 
\bibitem{burkhardt82} {Burkhardt T W 1982}\ {\PRL}\ {\bf 48}\ {216}
\bibitem{burkhardt82b} {Burkhardt T W 1982}\ {\PR\ \it B}\ {\bf
25}\ {7048}
\bibitem{turban99} {Turban L 1999}\ {\it Eur. Phys. J. B}\ {\bf **}\ 
{**}
\bibitem{kogut79} {Kogut J 1979}\ {\RMP}\ {\bf 51}\ {659} 
\bibitem{lieb61} {Lieb E H, Schultz T D and Mattis D C
1961}\ {\APNY}\ {\bf 16}\ {406}
\bibitem{peschel84} {Peschel I 1984}\ {\PR\ \it B}\ {\bf 30}\ {6783}
\bibitem{henkel88}{Henkel M and Sch\"utz G 1988}\ {\JPA}\ {\bf 21}\
{2617}
\bibitem{sornette98} {Sornette D 1998}\ {\it Phys. Rep.}\ {\bf 297}\
{239} 
\bibitem{igloi93} {Igl\'oi F, Peschel I and Turban L 1993}\ {\it Adv.  
Phys.}\ {\bf 42}\ {683}
\bibitem{harris74} {Harris A B 1974}\ {\JPC}\ {\bf 7}\ {1671} 
\bibitem{wu82} {Wu F Y 1982}\ {\RMP}\ {\bf 54}\ {235}
\bibitem{baxter73} {Baxter R J 1973}\ {\JPC}\ {\bf 6}\ {L445}
\bibitem{baxter82} {Baxter R J 1982}\ {\JPA}\ {\bf 15}\ {3329}
\bibitem{lipowsky82} {Lipowsky R 1982}\ {\PRL}\ {\bf 49}\ {1575}
\bibitem{igloi99b} {Igl\'oi F and Carlon E 1999}\ {\PR\ \it B}\ {\bf 
59}\
{3783}
\bibitem{imry79} {Imry Y and Wortis M 1979}\ {\PR\ \it B}\ {\bf 19}\
{3580} \bibitem{aizenman89} {Aizenman M and Wehr J 1989}\ {\PRL}\ {\bf 
62}\
{2503}
\bibitem{hui89} {Hui K and Berker A N 1989}\ {\PRL}\ {\bf 62}\ {2507}
\bibitem{nienhuis75} {Nienhuis B and Nauenberg M 1975}\ {\PRL}\ {\bf
35}\ {477}
\bibitem{fisher82} {Fisher M E and Berker A N 1982}\ {\PR\ \it B}\ {\bf 
26}\
{2507}

\vfill\eject\bye